\documentclass[fleqn,usenatbib]{mnras}

\usepackage[T1]{fontenc}

\DeclareRobustCommand{\VAN}[3]{#2}
\let\VANthebibliography\thebibliography
\def\thebibliography{\DeclareRobustCommand{\VAN}[3]{##3}\VANthebibliography}

\usepackage{graphicx}	% Including figure files
\usepackage{amsmath}	% Advanced maths commands
\usepackage{amssymb}	% Extra maths symbols
\usepackage{subfigure}  % added by author 
\usepackage{caption}    % added by author
\usepackage{hyperref}   % added by author

\usepackage{newtxtext, newtxmath}
\title[Dynamical evolution with INDICATE]{The dynamical evolution of star-forming regions measured with INDICATE}

\author[Blaylock-Squibbs \& Parker]{
George A. Blaylock-Squibbs\thanks{Email: gablaylock-squibbs1@sheffield.ac.uk} and
Richard J. Parker\thanks{Email: r.parker@sheffield.ac.uk}\thanks{Royal Society Dorothy Hodgkin fellow}
\\
Department of Physics and Astronomy, The University of Sheffield, Hounsfield Road, Sheffield, S3 7RH\\
}

\date{Accepted XXX. Received YYY; in original form ZZZ}

\pubyear{2023}

\begin{document}
\label{firstpage}
\pagerange{\pageref{firstpage}--\pageref{lastpage}}
\maketitle

\begin{abstract}
Observations of star-forming regions provide snapshots in time of the star formation process, and can be compared with simulation data to constrain the initial conditions of star formation. In order to make robust inferences, different metrics must be used to quantify the spatial and kinematic distributions of stars. In this paper, we assess the suitability of the INDICATE (\textbf{IN}dex to \textbf{D}efine
\textbf{I}nherent \textbf{C}lustering \textbf{A}nd \textbf{TE}ndencies) method as a diagnostic to infer the initial conditions of star-forming regions that subsequently undergo dynamical evolution. We use INDICATE to measure the degree of clustering in $N$-body simulations of the evolution of star-forming regions with different initial conditions. We find that the clustering of individual stars, as measured by INDICATE, becomes significantly higher in simulations with higher initial stellar densities, and is higher in subvirial star-forming regions where significant amounts of dynamical mixing has occurred.  We then combine INDICATE with other methods that measure the mass segregation ($\Lambda_{\rm{MSR}}$), relative stellar surface density ratio ($\Sigma_{\rm{LDR}}$) and the morphology (Q-parameter) of star-forming regions, and show that the diagnostic capability of INDICATE increases when combined with these other metrics.   
\end{abstract}


%%%%%%%%%%%%%%%%%%%%%%%%%%%%%%%%%%%%%%%%%%%%%%%%%%

%%%%%%%%%%%%%%%%% BODY OF PAPER %%%%%%%%%%%%%%%%%%

\section{Introduction}
Most stars form in groupings of 10s to 1000s of members, located along dense filaments embedded within giant molecular clouds (GMCs) \citep{lada_embedded_2003, andre_filamentary_2014}. By understanding the origins of star-forming regions\footnote{In this paper we use the term `star-forming region' to refer to a population of young stars that we assume formed from the same Giant Molecular Cloud. This star-forming region may be gravitationally bound, in which case it would be classified as a cluster, or unbound, in which case it would be an association. As we model the evolution of both bound and unbound stellar populations in this paper, we use `star-forming region', though some researchers exclusively reserve this term for populations of stars still surrounded by gas left over from the star formation process.} and their subsequent evolution and dispersal we can gain further understanding of Galactic evolution and the effects on planetary systems \citep{laughlin1998,smith2001,bonnell2001,parker_quanz2012,daffern-powell_great_2022,rickman2023}. 

As an example, a planetary system is unlikely to suffer dynamical perturbations in a low-density star-forming region that rapidly disperses into the Galactic field, compared to a high-density region that remains gravitationally bound for many dynamical timescales \citep{vincke_cluster_2016}. In the latter case, the orbits of planets can be altered, either through direct interactions \citep[e.g.][]{parker_quanz2012,daffern-powell_great_2022}, or subsequently, via secondary effects such as von Zeipel-Lidov-Kozai cycles \citep{fabrycky2007,malmberg2007}.

Recent work has suggested that even low-density star-forming regions are detrimental to planet formation if massive stars are present, whose FUV and EUV radiation fields are strong enough to evaporate protoplanetary discs during formation \citep{scally_destruction_2001,adams2004,winter2018,concha-ramirez_external_2019,haworth2023}.  

We therefore need to quantify the evolution of star-forming regions so that we can quantify the effects of the star-forming environment on planetary systems, by determining how long a given star-forming region is likely to remain gravitationally bound before dispersing into the Galactic field.% due to \textbf{gas expulsion or exhaustion if a much larger portion of gas ends up as stars \citep{kruijssen_fraction_2012}.} 

However, an observation of one star-forming region only provides information about the stellar density and velocity field at that snapshot in time. If all star-forming regions formed with the same initial conditions, we could build up a statistical picture of the evolution of star-forming regions by observing more than one region. 

This approach is complicated by the likelihood that star-forming regions form with a range of masses \citep{portegies_zwart_2010}, a range of stellar densities \citep{parker_dynamics_2014,parker_dynamical_2017,parker2022}, different degrees of initial substructure \citep{girichidis2012, dale2012,dale2013, dale2014, dib_and_henning_sf_spatial_distribution_ms_2019, daffern-powell_dynamical_2020, dib_et_al_characteristics_mcs_2020, schneider_understanding_sf_2022} and a range of initial virial states, with compact, bound clusters being the tail of a broad distribution \citep{kruijssen_fraction_2012}. This means that two star-forming regions that had very different initial conditions may evolve to be similar in appearance, or vice versa, something termed the `density degeneracy problem'. 

To overcome this problem, and to pinpoint the initial conditions of an observed star-forming region, different measures of the spatial distribution of stars, e.g.\,\,the mean surface density of companions \citep{larson_star_1995,bate_interpreting_1998,gouliermis_complex_2014}, the Q-parameter \citep{cartwright_statistical_2004, schmeja_et_al_structure_q_2008, sanchez_spatial_2009, lomax_statistical_2011, dib_et_al_structure_ms_stellar_clusters_2018, dib_and_henning_sf_spatial_distribution_ms_2019} and the spatial distribution of massive stars relative to the low-mass stars \citep[e.g.][]{allison_using_2009,kupper_mass_2011,maschberger_global_2011}, can be used to provide information. 

Previous work has shown that combinations of these measures can be used as a dynamical clock, to infer the amount of dynamical evolution that a star-forming region has undergone \citep[e.g][]{parker_dynamical_2014_mnras,parker_dynamics_2014,parker_dynamical_2017,blaylock-squibbs_evolution_2023}. Whilst these combinations are powerful diagnostics, they can be degenerate, especially for low-density star-forming regions, and new measures of clustering are continuously being added to the literature that require extensive testing against more established metrics. 
  
%A new method to measure clustering called INDICATE (\textbf{IN}dex to \textbf{D}efine
% \textbf{I}nherent \textbf{C}lustering \textbf{A}nd \textbf{TE}ndencies) has been tested on synthetic data, observations and SPH simulations \citep{buckner_spatial_2019, nony_mass_2021, buckner_+_2d_perspective_effects_2022, blaylock-squibbs_investigating_2022}. 

First introduced by \citet{buckner_spatial_2019} INDICATE (\textbf{IN}dex to \textbf{D}efine \textbf{I}nherent \textbf{C}lustering \textbf{A}nd \textbf{TE}ndencies) is designed to quantify the relative clustering of stars. INDICATE has been used to investigate NGC 3372 and NGC 2264, and also for assessing how accurately \textit{Gaia} can observe young stellar clusters and associations \citep{buckner_spatial_2019, nony_mass_2021, buckner+_can_gaia_accurately_observe_young_clusters_and_associations_2023}. Until now INDICATE has not been applied to purely $N$-body simulations of the long-term (10 Myr) dynamical evolution of star-forming region.

In this paper we measure the clustering evolution of $N$-body simulations of star-forming regions using INDICATE. We also combine INDICATE with other methods (the Q-parameter \citep{cartwright_statistical_2004}, $\Sigma_{\rm{LDR}}$ \citep{maschberger_global_2011}, $\Lambda_{\rm{MSR}}$ \citep{allison_using_2009}) for different snapshots in our simulations to assess the diagnostic ability of INDICATE to pinpoint the initial conditions of star-forming regions.

The paper is organised as follows. In Section~2 we describe the set-up of our $N$-body simulations, and briefly define the different metrics for quantifying the spatial distribution of stars within the region, including INDICATE. In Section 3 we present our results, and we conclude in Section~4.

\section{Methods}
In this section we describe the setup of the $N$-body simulations before describing the methods used to quantify the clustering, morphology, surface density and mass segregation of star-forming regions as they dynamically evolve.

\subsection{$N$-Body simulation set up}
We utilise the simulations previously described in \cite{blaylock-squibbs_evolution_2023}. We have eight sets of simulations, each with different initial conditions (i.e. initial degree of substructure, density and virial state). We run 10 regions for each set of initial conditions, as even though they share the same initial properties, there is some stochasticity in the dynamical evolution and two statistically similar star-forming regions can evolve very differently from one another \citep{parker_dynamical_2014_mnras}. 
 
% %%%%%%%%%%%%%%%%%%%%%%%%%%%%%%%%%
% The simulations are run using \texttt{Kira} which is a part of the \texttt{Starlab} package \citep{zwart_star_1999, portegies_zwart_star_2001}. The simulations contain 1000 stars, this choice is motivated by the following observed relation for young embedded star clusters:
% \begin{equation}
%     N_{\rm{cl}} \propto M_{\rm{cl}}^{-2},
% \end{equation}
% where $N_{\rm{cl}}$ is the number of stellar clusters and $M_{\rm{cl}}$ is the mass of a cluster \citep{lada_embedded_2003}. This relation is observed for cluster masses of $10 < M_{\rm{cl}} / M_{\odot} < 10^{5}$, which places our simulated \textbf{star-forming regions} of 1000 stars near to the middle of the distribution (see masses in Table~\ref{table:simulation initial conditions}).
% %%%%%%%%%%%%%%%%%%%%%%%%%%%%%%%%%
Our simulated regions contain 1000 stars, with average total masses of $\sim600\,$M$_{\odot}$, which places them in the middle of the cluster mass distribution from \citet{lada_embedded_2003}, which ranges from $10\,$M$_{\odot}$ to $10^{5}\,$M$_{\odot}$.

We set up the simulations with two very different velocity fields, as defined by the virial ratio $\alpha_{\rm vir} = T/|\Omega$|, where $T$ and |$\Omega$| are the total respective kinetic and potential energies of the stars. Observations of prestellar cores show them to have a subvirial velocity dispersion (main sequence stars will inherit their velocities from the prestellar cores) \citep{foster_-sync_2015, kuznetsova_signatures_2015}, and so we run a set of subvirial simulations ($\alpha_{\rm vir} = 0.1$). We also run sets of supervirial simulations ($\alpha_{\rm vir} = 0.9$) as the observations of \citet{bravi_gaia-eso_2018, kuhn_kinematics_2019} and \citet{kounkel_dynamical_2022} show that some young (around 1-5 Myr) star-forming regions are expanding.

\begin{table}
	\centering
	\caption{Table summarising the initial conditions for the eight sets of 10 simulations. From left to right the columns show the fractal dimension (lower values correspond to greater degrees of substructure), the initial radii of the simulations, the median initial stellar mass density, the mean total stellar mass in the sets and the initial virial state of the regions where they can either be collapsing or expanding, $\alpha_{\rm{vir}} = 0.1$ or $\alpha_{\rm{vir}} = 0.9$, respectively.}
	\label{table:simulation initial conditions}
	\begin{tabular}{lccccc} % four columns, alignment for each
		\hline
		$D_{f}$ & r (pc) & $\tilde{\rho}$ (M$_{\odot}\,$pc$^{-3}$) & $\bar{\rm{M}} (\rm{M}_{{\odot}})$ & $\alpha_{\rm{vir}}$ & $N_{\star}$\\
		\hline
		1.6 & 1 & $10^{4}$  & 592 & 0.1   & 1000 \\
		1.6 & 1 & $10^{4}$  & 592 & 0.9   & 1000 \\
		1.6 & 5 & $10^{2}$  & 592 & 0.1 & 1000 \\
        1.6 & 5 & $10^{2}$  & 592 & 0.9 & 1000 \\
        3.0 & 1 & $10^{2}$  & 624 & 0.1   & 1000 \\
        3.0 & 1 & $10^{2}$  & 624 & 0.9   & 1000 \\
        3.0 & 5 & $10^{0}$  & 624 & 0.1 & 1000 \\
        3.0 & 5 & $10^{0}$  & 624 & 0.9 & 1000 \\
		\hline
	\end{tabular}
\end{table}

\subsubsection{Substructure}

Observations of young star-forming regions show that stars appear to form in filamentary structures, with young stars exhibiting spatial and kinematic substructure \citep{efremov_+_hierarchical_star_formation_1998, andre_filamentary_2014, plunkett_et_al_distribution_of_stars_serpens_2018, dib_and_henning_sf_spatial_distribution_ms_2019, hacar_initial_2022}.

We model this substructure in our simulations using the box-fractal method, generating simulations with a high degree of substructure (corresponding fractal dimension of $D_{f} = 1.6$) and simulations with no substructure (corresponding fractal dimension of $D_{f} = 3.0$) \citep{goodwin_dynamical_2004, daffern-powell_dynamical_2020}.

To generate substructure we follow the method of \citet{cartwright_statistical_2004, goodwin_dynamical_2004} which has been used extensively in the literature \citep{allison_using_2009, parker_comparisons_2015, daffern-powell_dynamical_2020, daffern-powell_great_2022, blaylock-squibbs_evolution_2023}. The box-fractal method works as follows. A single star is placed at the centre of a cube with side length $N_{\rm{Div}} = 2$ \citep[in order to create substructure, $N_{\rm{Div}}$ must be greater than unity, but the choice of $N_{\rm{Div}} = 2$ is arbitrary, see][]{goodwin_dynamical_2004}. This cube is then subdivided into $N_{\rm{Div}}^3$ (8 in this case) smaller sub-cubes. A star is then placed at the centre of each of the sub-cubes. Each sub-cube then has a probability of $N_{\rm{Div}}^{D_{f}-3}$ of being subdivided itself, where $D_{f}$ is the desired fractal dimension of the region. The lower the fractal dimension, $D_{f}$, the more substructured the region will be. For example, if $D_{f} = 1.6$, then the probability of that star's cube being subdivided is $N_{\rm{Div}}^{-1.4}$, whereas for $D_{f} = 3.0$, then the probability of that star's cube being subdivided is $N_{\rm{Div}}^{0}$, i.e.\,\,unity. 

Stars whose cubes are not subdivided are removed, along with any previous generation of stars that preceded them. A small amount of random noise is added to the position of the stars to remove any regular structure that may appear. The stars' cubes are subdivided repeatedly until the desired number of stars ($N_{\star} = 1000$) is reached or exceeded. Once a generation consists of or exceeds the target number of stars, all previous generations of stars are removed. In the event the target number of stars is exceeded, we randomly select stars in the last generation and remove them until the target number of stars is reached.

\subsubsection{Stellar Velocities}
The initial star in the box-fractal method has a velocity drawn from a Gaussian with mean 0 km\,s$^{-1}$ and variance 1 km\,s$^{-1}$. Subsequent stars inherit this velocity plus a value drawn from the same Gaussian but multiplied by a factor of $(1 / N_{\rm{Div}})^{g}$, where $g$ is the current generation of stars. Because of this, the velocity inherited decreases for each generation which results in stars spatially close to one another having similar velocities whereas stars that are far apart spatially can have very different velocities. This means that our simulations also have kinematic substructure, which may be able to trace the formation mode of star clusters \citep{arnold_quantifying_2022}.

We then scale the velocities of the stars to the desired initial virial ratio of the simulations, where the virial ratio $\alpha_{\rm vir} = T/|\Omega|$, where $T$ and $\Omega$ are the total kinetic and potential energies of the stars, respectively, and $\alpha_{\rm vir} = 0.5$ is virial equilibrium. We set the simulations to be  either subvirial (cool, collapsing regions, $\alpha_{\rm vir} = 0.1$) or supervirial (warm, expanding regions, $\alpha_{\rm vir} = 0.9$).

\subsubsection{Stellar masses}
Stellar masses in the simulations are drawn randomly from the Maschberger IMF \citep{maschberger_function_2013}. The Maschberger IMF has a functional form, unlike the piece-wise IMFs from \citet{kroupa_variation_2001} and \citet{chabrier_galactic_2003}. The probability density function of the Maschberger IMF is 
\begin{equation}
    p(m) \propto \left(  \frac{m}{\mu}  \right)^{-\alpha}  \left(  1 + \left(  \frac{m}{\mu}  \right)^{1-\alpha}  \right)^{-\beta},
\end{equation}
where $\alpha = 2.3$ is the high mass exponent and $\beta = 1.4$ describes the slope of the IMF for lower masses. $\mu = 0.2\,$M$_\odot$ is the scale factor and the combination of these parameters produces a mean stellar mass of $\sim 0.4$\,M$_\odot$. This form of the IMF is very similar to other parameterisations \citep[e.g.][]{kroupa_dense_2008, chabrier_galactic_2003}. We note that other versions, where parameters such as the exponent $\alpha$ may vary with the surface density of stars, have been proposed in the literature \citep{dib_imf_paper_2023}.

The masses are selected using the quantile function
\begin{equation}
    m(u) = \mu \left(   \left[  u \left(G(m_{\rm upp}) - G(m_{\rm low})\right) + G(m_{\rm low}) \right]^{\frac{1}{1-\beta}}  - 1  \right)^{\frac{1}{1-\alpha}},
\end{equation}
where the input $u$ is a value drawn from a uniform distribution between $0 < u < 1$, $m_{\rm low} = 0.1\,$M$_{\odot}$ is the lower mass limit and $m_{\rm upp} = 50\,$M$_{\odot}$ is the upper mass limit.

The function $G(m)$ is the auxiliary function
\begin{equation}
    G(m) = \left(  1 + \left(   \frac{m}{\mu}    \right)^{1-\alpha}   \right)^{1-\beta},
\end{equation}
where $m$ is either $m_{\rm low}$ or $m_{\rm upp}$ and the other the terms are as above.

\subsubsection{Dynamical evolution}
We take the masses, positions and velocities of the stars and evolve them using the 4$^{\rm th}$-order Hermite scheme \texttt{kira} integrator within the \texttt{Starlab} software environment \citep{portegies_zwart_star_2001}. The simulations are evolved for 10\,Myr and we do not include stellar evolution, nor do we simulate the background gas potential in our simulations nor the galactic tidal field. Both of these likely affect the dispersal of young stellar clusters via gas expulsion or by tidal stripping, reducing the time simulations remain bound \citep{fellhauer_and_Kroupa_star_cluster_survival_2005, Mamikonyan+_evolution_of_star_clusters_in_tidal_fields_2017}. \\

A summary of the initial conditions for the $N$-body simulations is given in Table~\ref{table:simulation initial conditions}.

\subsection{INDICATE}
\label{methods:subsec:indicate}
To measure the clustering of stars in our simulations we use INDICATE \citep{buckner_spatial_2019}. INDICATE measures the degree of relative clustering on a star by star basis and also determines the significance of any clustering. %\textbf{INDICATE has the advantage that no prior assumptions about the regions are needed for the method to work.}

The INDICATE algorithm proceeds as follows. First, an evenly spaced control grid is generated with the same number density as the data of interest. This control grid has the appearance of a regular grid-like distribution of points that extends beyond the original data set. Because the control grid is constructed over the same spatial scale, the degree of clustering measured by INDICATE is already normalised, allowing the degree of clustering measured by INDICATE to be compared with the clustering in different regions, without needing to normalise the datasets against each other \citep[the Q-parameter][is normalised in a similar way]{cartwright_statistical_2004}. The number density of the data is calculated by dividing the number of stars by the rectangular area that encloses all stars. The rectangle is defined using the minimum and maximum $(x,y)$ coordinates of stars in the region. Then, for each star we calculate the Euclidean distance to its $N^{\rm{th}}$ nearest neighbour in the control grid. The mean of these distances is then found, which we call $\bar{r}$. For each star, $j$, we calculate the number of other stars within $\bar{r}$ of $j$. The INDICATE index is 
\begin{equation}
    I_{j, N} = \frac{N_{\bar{\rm{r}}}}{N},
\end{equation}
where $N_{\bar{\rm{r}}}$ is the number of stars within $\bar{r}$ of star $j$ and $N$ is the nearest neighbour number (we follow \citet{buckner_spatial_2019} and use $N = 5$ in this work). In this work we characterise the overall clustering in the simulations at each snapshot by calculating the mean INDICATE index, $\bar{I}_{5}$.

\subsubsection{Significant Index}
INDICATE can determine the significance of clustering on a star-by-star basis by using a significant index, above which stars are non-randomly clustered, and is calculated as follows. Firstly, once the control grid has been generated for the data we then generate a uniform distribution of points with the same number density as the data. We then calculate the INDICATE indices for the uniform distribution of points and calculate the mean of all of these indices. The significant index is defined as 
\begin{equation}
    I_{\rm{sig}} = \bar{I} + 3 \, \sigma,
\end{equation}
where $\bar{I}$ is the mean INDICATE index found for the uniform distribution of points and $\sigma$ is the standard deviation of the indices for the uniform distribution. \citet{buckner_spatial_2020}  calculate the significant index by repeating the significant index calculation 100 times, each time using a different uniform distribution, then finding the mean significant index. \citet{blaylock-squibbs_investigating_2022} show that for $\sim1000$ stars a single calculation is sufficient, giving a similar significant index ($\sim2.2$) to the one calculated using repeats. However, for data sets with < 100 stars a minimum of 20 repeats should be used.

\subsection{Q-Parameter}
The Q-parameter was first introduced in \citet{cartwright_statistical_2004} and is used to differentiate between regions with different morphologies. The Q-parameter uses a minimum spanning tree (MST), which is a graph connecting all points in such a way that the total edge length of the graph is minimised, and there are no closed loops.

The Q-parameter is a ratio between the normalised mean edge length, $\bar{m}$, and the normalised correlation length, $\bar{s}$, and is calculated as follows. First, the normalised mean edge length is calculated by generating an MST of all stars in the data and finding the mean edge length of this MST. The mean edge length is normalised by dividing it by $\frac{\sqrt{N_{\rm{total}} \, A}}{N_{\rm{total}} - 1}$, where $A$ is the circular area of the region (though see \citet{schmeja_evolving_2006} and \citet{parker_spatial_2018} for a discussion on alternative normalisation methods).

The normalised correlation length is the mean separation between all stars, which is normalised by dividing the mean separation by the radius of the region. The Q-parameter is defined as
\begin{equation}
    Q = \frac{\bar{m}}{\bar{s}}.
\end{equation}

The value of the Q-parameter indicates the morphology of a group of stars; $Q < 0.8$ implies a substructured region, whereas $Q > 0.8$ implies a  a smooth, centrally concentrated region.% with 

\subsection{Local stellar surface density ratio: $\Sigma_{\rm LDR}$}
The local stellar surface density ratio was first used in \citet{maschberger_global_2011} to quantify the differences in stellar surface density of chosen subsets of stars compared to the entire region. In this work our chosen subset is the 10 most massive stars. The method proceeds as follows. First, for each star we calculate the two dimensional euclidean distance to the $N^{th}$ nearest neighbour, we use $N = 5$ in this work. The surface density of a star is then defined as
\begin{equation}
    \Sigma = \frac{N - 1}{\pi \, R^{2}_{N}},
\end{equation}
where $N = 5$ and $R$ is the distance from the star to its the $5^{th}$ nearest neighbour \citep{casertano_core_1985}. The local surface density ratio is defined as
\begin{equation}
    \Sigma_{\rm{LDR}} = \frac{\Tilde{\Sigma}_{\rm{subset}}}{\Tilde{\Sigma}_{\rm{all}}},
\end{equation}
where $\Tilde{\Sigma}_{\rm{subset}}$ is the median surface density of the 10 most massive stars and $\Tilde{\Sigma}_{\rm{all}}$ is the median surface density of all stars in our simulations. If $\Sigma_{\rm{LDR}} > 1$ then the 10 most massive stars are located in areas of higher than average surface density, and if $\Sigma_{\rm{LDR}} < 1$ they are located in areas of lower than average surface density. We quantify the significance of any deviation from unity via a two-sample Kolmogorov-Smirnov (KS) test, where a $p-$value of less than 0.01 is associated with the difference between the subset of massive stars, and the entire sample, means we reject the hypothesis that they share the same underlying parent distribution. Projection effects do not unduly affect this measurement \citep{bressert_spatial_2010,parker_characterizing_2012}.

\subsection{Mass segregation ratio: $\Lambda_{\rm MSR}$}
The mass segregation ratio is a metric to quantify the degree of mass segregation in a star-forming region, and was first developed by \citet{allison_using_2009}, and has been used extensively \citep{olczak_et_al_mass_seg_method_2011, parker_dynamical_2017, dib_et_al_structure_ms_stellar_clusters_2018, dib_and_henning_sf_spatial_distribution_ms_2019, plunkett_et_al_distribution_of_stars_serpens_2018, maurya_et_al_stats_of_dyn_evolution_open_clusters_2023}. Mass segregation is when the separation  between the most massive stars is smaller than the separation between the average stars in a region (for example, if the massive stars are all located in the centre of a star-forming region). The method proceeds as follows. First we generate an MST for a chosen subset of stars; in this work we select the 10 most massive stars. We then generate MSTs for 10 randomly chosen stars (which can include stars from the chosen subset); we repeat this 200 times and calculate the mean edge length for the random MSTs. The mass segregation ratio is
\begin{equation}
    \Lambda_{\rm{MSR}} = \frac{\left<l_{\rm{average}}\right>}{l_{10}}^{+ \sigma_{5/6} / l_{10}}_{- \sigma_{1/6} / l_{10}},
\end{equation}
where $\left<l_{\rm{average}}\right>$ is the mean edge length for the randomly constructed MSTs and $l_{10}$ is the edge length of the MST for the 10 most massive stars. We follow \citet{parker_spatial_2018} and find the upper ($+ \sigma_{5/6} / l_{10}$) and lower ($- \sigma_{1/6} / l_{10}$) uncertainties by taking an ordered list of all of the random MST lengths and selecting the upper and lower uncertainties from 5/6 and 1/6 of the way through the ordered list of MST lengths. The uncertainties will therefore corresponds to a 66\% deviation from the median MST length and prevents outliers from affecting the uncertainty, which could be an issue if a Gaussian dispersion is used to estimate the uncertainty \citep{allison_using_2009}.

In addition to the uncertainties, we follow \citet{parker_comparisons_2015} and require $\Lambda_{\rm{MSR}} > 2$ as a significant detection of mass segregation. Therefore if $\Lambda_{\rm {MSR}} > 2$ then the 10 most massive stars are said to be mass segregated (the 10 most massive stars are closer to each other than the average stars are to each other) and if $\Lambda_{\rm {MSR}} \sim 1$ then they are not mass segregated (the 10 most massive stars are separated by a similar distance to the average stars). If $\Lambda_{\rm{MSR}} < 1$ the region is said to be inversely mass segregated, with the most massive stars more widely distributed compared to the average stars in the region.

\section{Results}
In this section we present our results in which we follow the evolution of the INDICATE metric, $I_{5}$, for $N$-body simulations with a high initial degree of substructure ($D_{f} = 1.6$) and simulations with little to no initial substructure ($D_{f} = 3.0$). We show results for both high density ($10^{2} - 10^{4}\,$M$_{\odot}\,$pc$^{-3}$) and low density ($10^{0} - 10^{2}\,$M$_{\odot}\,$pc$^{-3}$) realisations of these simulations, corresponding to initial radii of 1 pc and 5 pc, respectively. We quantify the clustering using INDICATE along three different lines of sight (LOS); each LOS being parallel to one of the component axes (i.e. $(x,y,z)$). 

We then present the results where we combine $Q$, $\Sigma_{\rm{LDR}}$ and $\Lambda_{\rm{MSR}}$ with INDICATE, and assess which combinations most reliably constrain the initial conditions of the simulated star-forming regions.

% 1d6 1pc radii
\begin{figure*}
 \subfigure[subvirial (x,y)]{\includegraphics[width=0.31\linewidth]{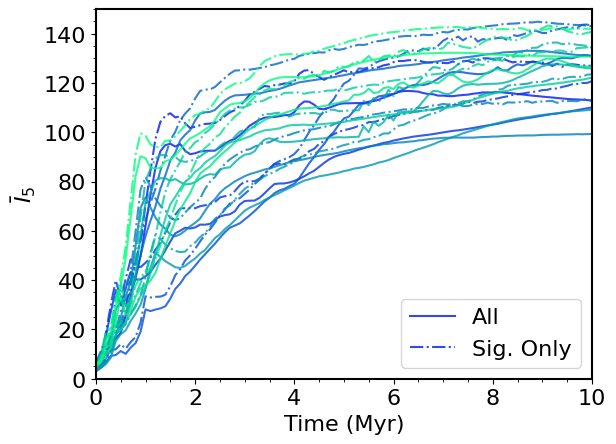}}
 \hspace{0.8pt}
 \subfigure[subvirial (x,z)]{\includegraphics[width=0.31\linewidth]{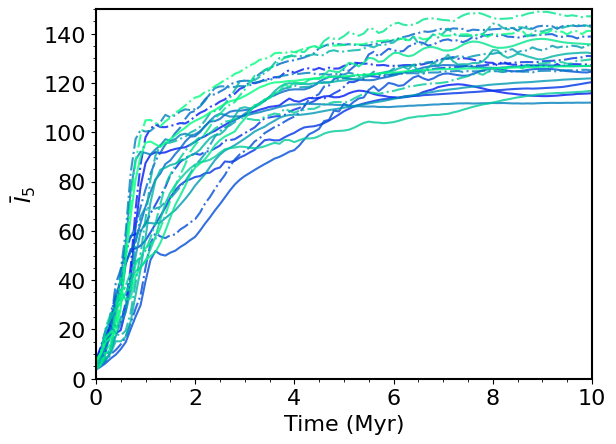}}
 \hspace{0.8pt}
 \subfigure[subvirial (y,z)]{\includegraphics[width=0.31\linewidth]{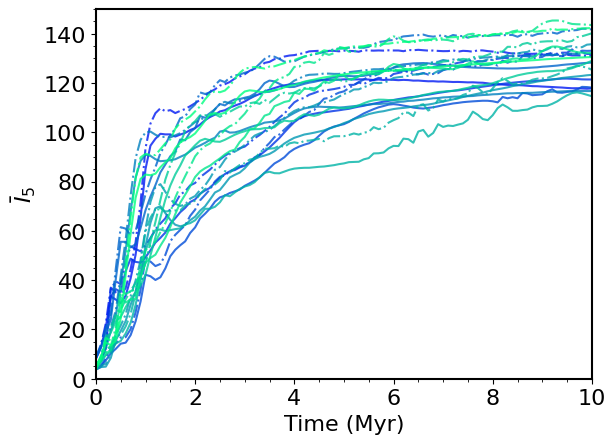}}
 \hspace{0.8pt}
 \subfigure[supervirial (x,y)]{\includegraphics[width=0.31\linewidth]{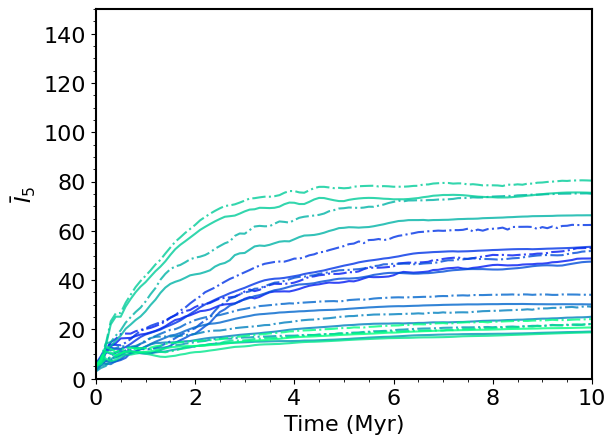}}
 \hspace{0.8pt}
 \subfigure[supervirial (x,z)]{\includegraphics[width=0.31\linewidth]{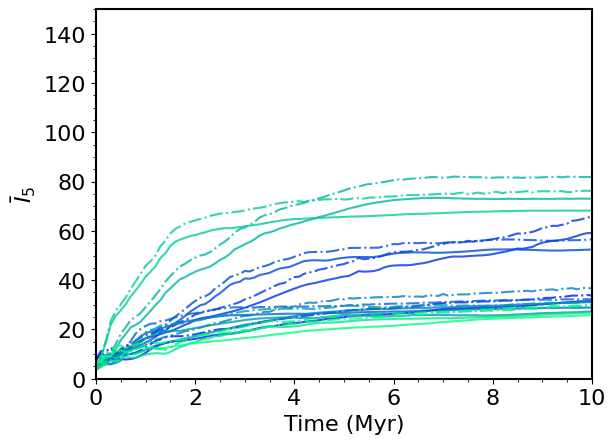}}
 \hspace{0.8pt}
 \subfigure[supervirial (y,z)]{\includegraphics[width=0.31\linewidth]{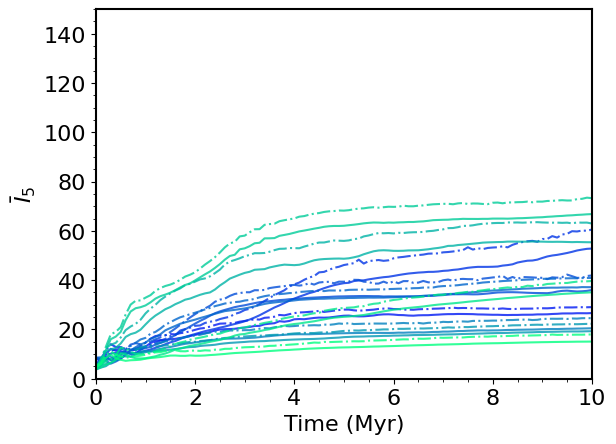}}
    
    \caption{The mean INDICATE index against time for high-density, substructured, sub- and supervirial simulations. Both the sub- and supervirial simulations have a high initial degree of substructure ($D_{f} = 1.6$) and both contain 1000 stars with initial radii of 1 pc, simulated for 10 Myr. From left to right the columns show the INDICATE indexes calculated along three different lines of sight, $(x,y)$, $(x,z)$ and $(y,z)$. The top row shows the indexes calculated for subvirial simulations and the bottom row shows the indexes calculated for supervirial simulations. The solid lines show the mean index for all stars and the dashed lines show the mean index for the significantly clustered stars. We show the 10 simulations we run for the different sets of initial conditions, with the colour shading of the lines being consistent across the different lines of sight.}
    \label{fig:1d6 1 pc}
\end{figure*}

\subsection{Evolution of INDICATE in substructured regions}
Figure~\ref{fig:1d6 1 pc} shows the mean INDICATE indexes for initially substructured simulations with fractal dimension $D_{f} = 1.6$ with initial radii of 1 pc. The general trend across all three lines of sight ($(x, y)$, $(x, z)$, $(y, z)$) for the subvirial regions (top row, panels a--c) is a rapid increase in the mean INDICATE index within the first 2 Myr, and then a gradual increase in the clustering for the remainder of the simulations. 

\begin{figure*}
\hspace*{-1.2cm}
 \subfigure[$(x,y)$]{\includegraphics[width=0.35\linewidth]{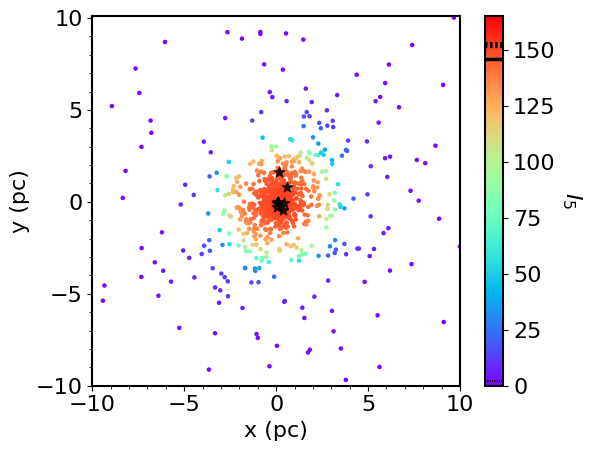}}
% \hspace{0.1pt}
 \subfigure[$(x,z)$]{\includegraphics[width=0.35\linewidth]{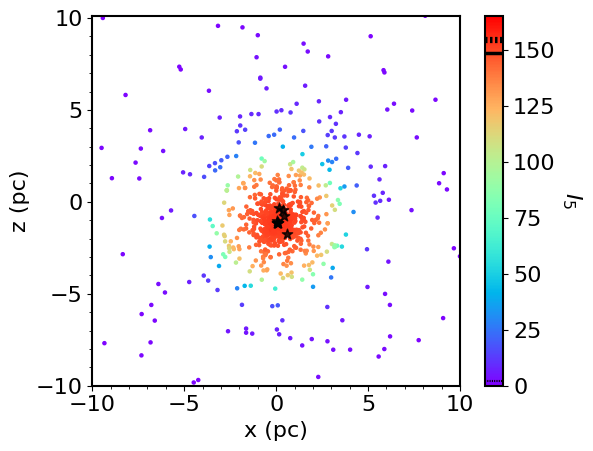}}
 %\hspace{0.1pt}
 \subfigure[$(y,z)$]{\includegraphics[width=0.35\linewidth]{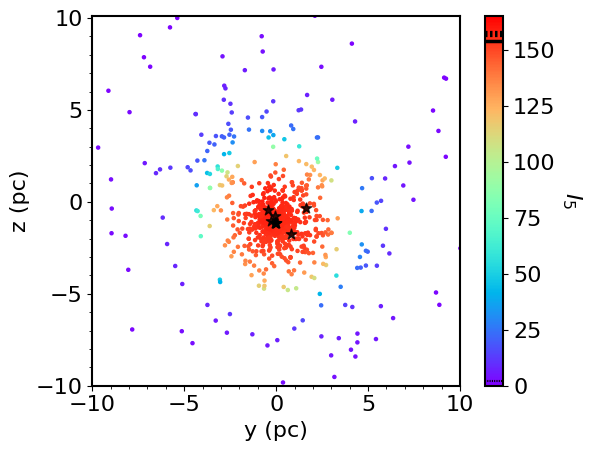}}

    \caption{Three different line-of-sight perspectives of a subvirial simulation of 1000 stars after 5 Myr of dynamical evolution. The panels show the simulation as viewed along the $z$, $y$ and $x$ axes. The colour of the points corresponds to their INDICATE index. The redder the points the more clustered a star is,  according to INDICATE. The 10 most massive stars are highlighted with star symbols. We show the median index for all 1000 stars with the solid black line in the colour bar ($\tilde{I}_{5,\,all} = 146.00, 148.60, 154.00$ calculated along the $z$, $y$ and $x$ axes, respectively), the black dashed line shows the median index for the 10 most massive stars ($\tilde{I}_{5, 10} = 152.10, 154.20, 156.90$ calculated along the same lines of sight) and the thin black dotted line shows the significant index ($I_{\rm{sig}} = 2.2$); a star with an index above this is said to be significantly clustered. The scale of the colour bar is scaled based on the maximum INDICATE index found in the $(y,z)$ plane of $I_{\rm{max}} = 158.20$.}
    \label{fig:cluster persepectives subvirial}
\end{figure*}

\begin{figure*}
\hspace*{-1.2cm}
 \subfigure[$(x,y)$]{\includegraphics[width=0.35\linewidth]{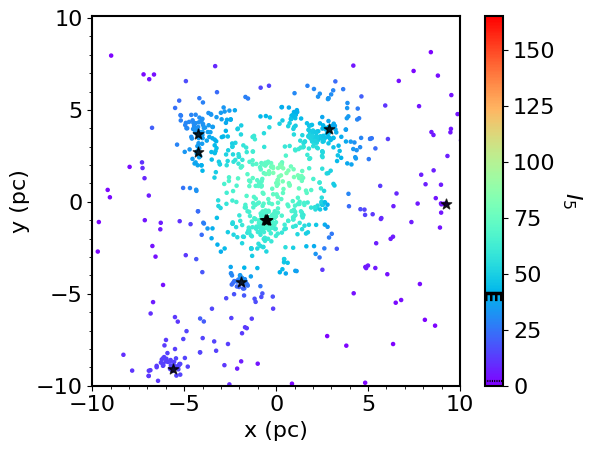}}
 %\hspace{0.8pt}
 \subfigure[$(x,z)$]{\includegraphics[width=0.35\linewidth]{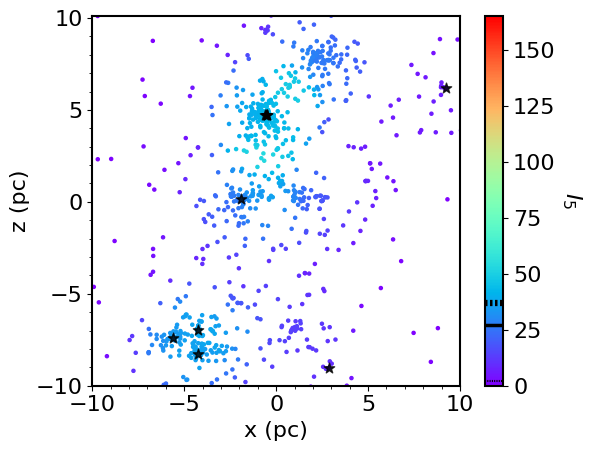}}
% \hspace{0.8pt}
 \subfigure[$(y,z)$]{\includegraphics[width=0.35\linewidth]{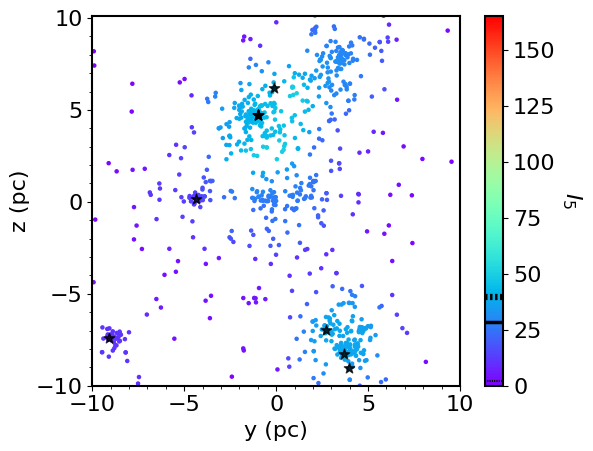}}

    \caption{Three different line-of-sight perspectives of a supervirial simulation of 1000 stars after 5 Myr of dynamical evolution. The panels show the simulation as viewed along the $z$, $y$ and $x$ axes. The colour of the points corresponds to their INDICATE index. The redder the points the more clustered a star is according to INDICATE. The 10 most massive stars are highlighted with star symbols. We show the median index for all 1000 stars ($\tilde{I}_{5,\,all} = 41.60, 27.00, 28.60$ calculated along the $z$, $y$ and $x$ axes, respectively) with the solid black line in the colour bar, the black dashed line shows the median index for the 10 most massive stars ($\tilde{I}_{5, 10} = 39.20, 37.00, 39.40$ calculated along the same lines of sight) and the thin dotted line shows the significant index ($I_{\rm{sig}} = 2.2$); a star with an index above this is said to be significantly clustered. The colour bar is scaled based on the maximum INDICATE index found in the $(y,z)$ plane for a the subvirial simulation (see Fig. \ref{fig:cluster persepectives subvirial}).}
    \label{fig:cluster persepectives supervirial}
\end{figure*}

We show how different initial conditions and perspectives affect the measured amount of clustering in simulations at 5 Myr along three different lines of sight in Figures~\ref{fig:cluster persepectives subvirial} and \ref{fig:cluster persepectives supervirial}. Although there is some variation depending on the line of sight, this is minimal compared to the differences between the sub- and supervirial simulations. The supervirial simulations have a much lower degree of clustering.

% ============== Figure showing the different perspectives of a supervirial highly substructure simulation at 5 Myr.
% 1d6 5 pc radii

The same behaviour is seen for some of the supervirial simulations (Fig~\ref{fig:1d6 1 pc}, panels d--f), but with a much slower initial increase in the mean indexes. The supervirial simulations also never reach the same degree of clustering according to INDICATE that the subvirial simulations do. The lower final $\bar{I}_{5}$ in the supervirial simulations is because the stars are all initially moving away from one another. Some stars assemble themselves in small bound groupings,  but the overall lower number of stars in these subgroups results in lower values of $\bar{I}_{5}$.

The main difference between the different initial virial states is in the final mean indexes, with subvirial simulations attaining mean indexes of $\gtrsim100$ and supervirial simulations not exceeding a mean index of $\sim90$ by 10 Myr. This behaviour is seen across all three different lines of sight. The subvirial simulations attain higher $\bar{I}_{5}$ due to all the stars falling into the gravitational potential well and forming a cluster \citep{allison_early_2010,parker_dynamical_2014_mnras}, which also explains the rapid increase of $\bar{I}_{5}$ during the first 2 Myr. 

\begin{figure*}
 \subfigure[subvirial (x,y)]{\includegraphics[width=0.31\linewidth]{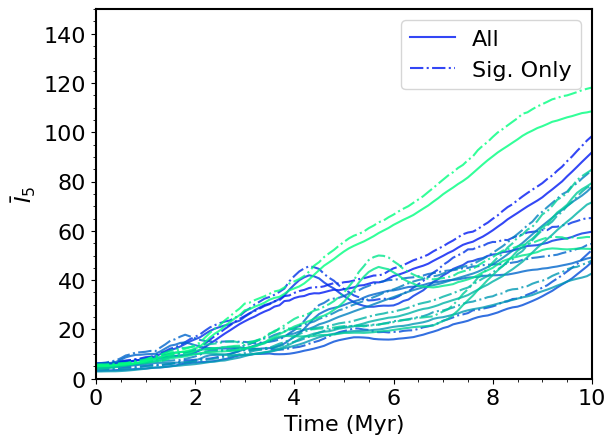}}
 \hspace{0.8pt}
 \subfigure[subvirial (x,z)]{\includegraphics[width=0.31\linewidth]{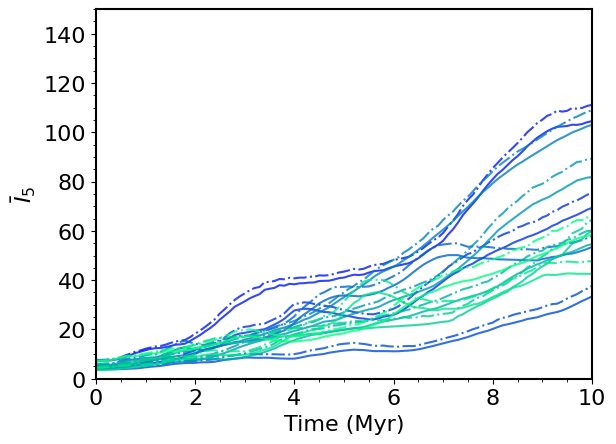}}
 \hspace{0.8pt}
 \subfigure[subvirial (y,z)]{\includegraphics[width=0.31\linewidth]{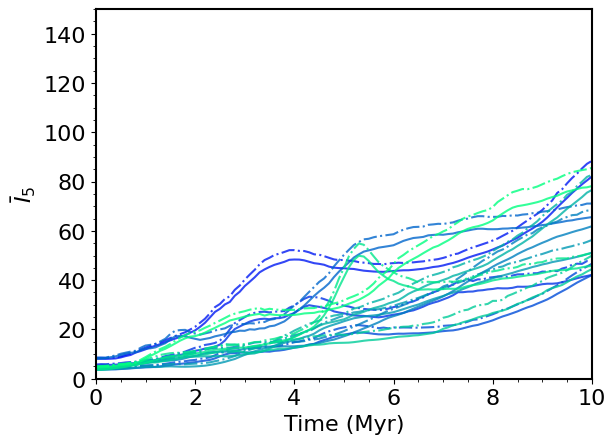}}
 \hspace{0.8pt}
 \subfigure[supervirial (x,y)]{\includegraphics[width=0.31\linewidth]{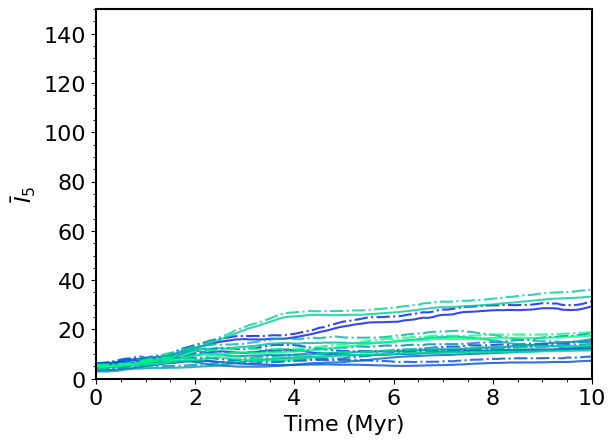}}
 \hspace{0.8pt}
 \subfigure[supervirial (x,z)]{\includegraphics[width=0.31\linewidth]{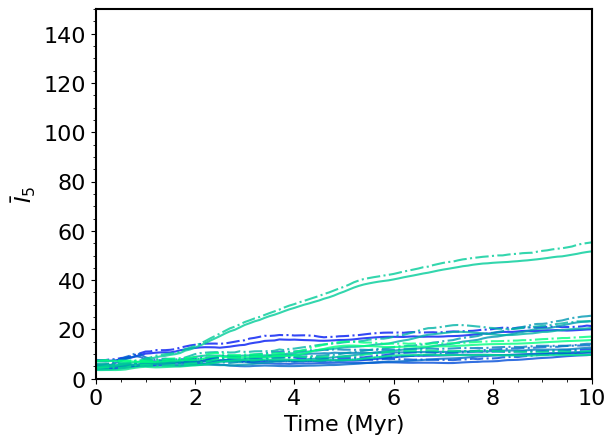}}
 \hspace{0.8pt}
 \subfigure[supervirial (y,z)]{\includegraphics[width=0.31\linewidth]{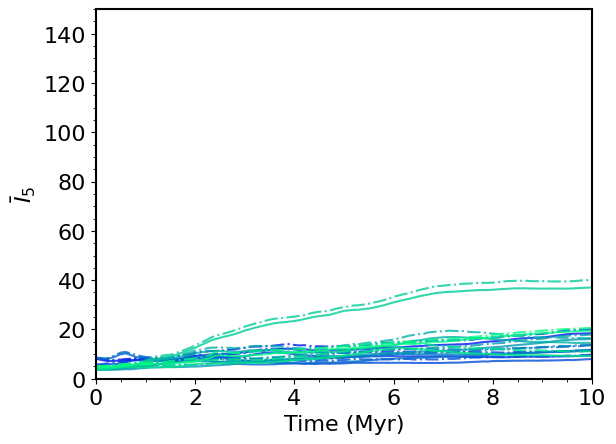}}
    
    \caption{The mean INDICATE index against time for low-density, substructured, sub- and supervirial simulations. Both the sub- and supervirial simulations have a high initial degree of substructure ($D_{f} = 1.6$) and both contain 1000 stars with initial radii of 5 pc, simulated for 10 Myr. From left to right the columns show the INDICATE indexes calculated for three different lines of sight, $(x,y)$, $(x,z)$ and $(y,z)$. The top row shows the indexes calculated for subvirial simulations and the bottom row shows the indexes calculated for supervirial simulations. The solid lines show the mean index for all stars and the dashed lines show the mean index for the significantly clustered stars. We show the 10 simulations we run for the different sets of initial conditions, with the colour shading of the lines being consistent across the different lines of sight.}
    \label{fig:1d6 5pc}
\end{figure*}

Figure~\ref{fig:1d6 5pc} shows the mean INDICATE indexes for substructured ($D_{f} = 1.6$) low-density simulations  with initial radii of 5 pc. We see the same general trend as in Figure~\ref{fig:1d6 1 pc} with the subvirial simulations attaining higher final mean indexes (panels a--c), compared to the supervirial simulations (panels d--f). A key difference is the lack of a rapid initial increase in the clustering; due to the low density of the simulations the stars take longer to interact and so the rate of clustering is slower. We see a maximum mean index of $\bar{I}_{5}\sim120$ for the subvirial simulations at 10 Myr, but this is only for one out of the 10 simulations. The other subvirial simulations finish with $40 \lesssim \bar{I}_{5} \lesssim 90$, below the minimum mean index found in the much denser ($r = 1\,$pc) subvirial simulations.

% 1d6 5 pc radii

\subsection{Evolution of INDICATE in smooth regions}
\begin{figure*}
 \subfigure[subvirial (x,y)]{\includegraphics[width=0.31\linewidth]{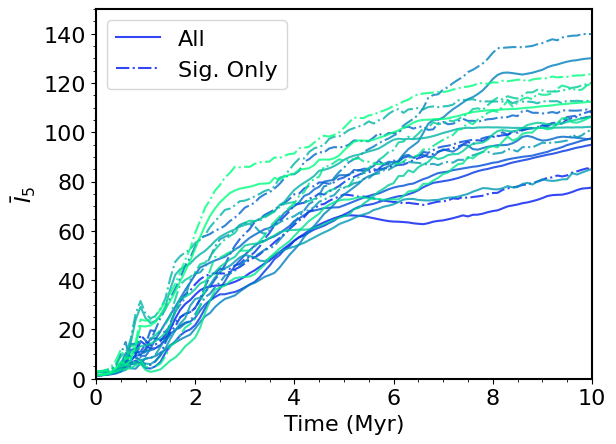}}
 \hspace{0.8pt}
 \subfigure[subvirial (x,z)]{\includegraphics[width=0.31\linewidth]{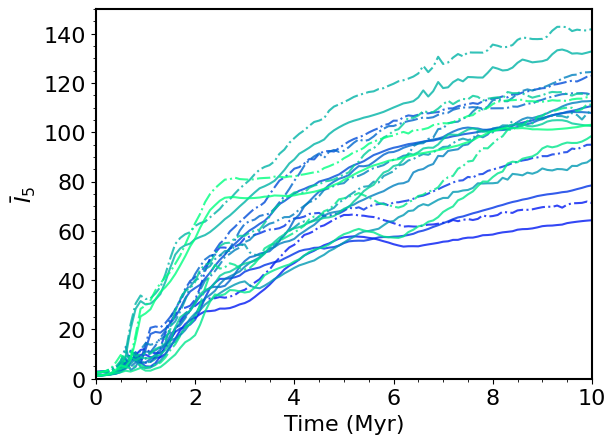}}
 \hspace{0.8pt}
 \subfigure[subvirial (y,z)]{\includegraphics[width=0.31\linewidth]{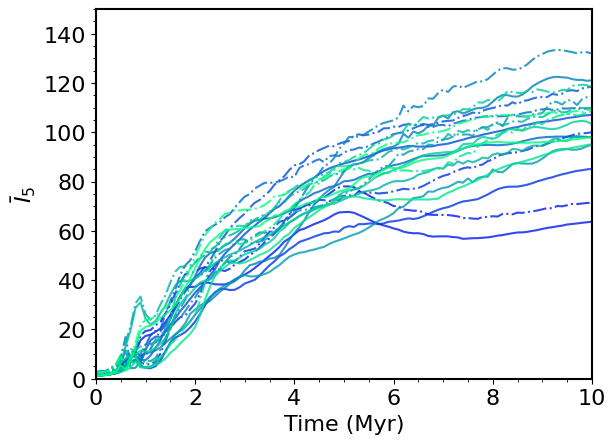}}
 \hspace{0.8pt}
 \subfigure[supervirial (x,y)]{\includegraphics[width=0.31\linewidth]{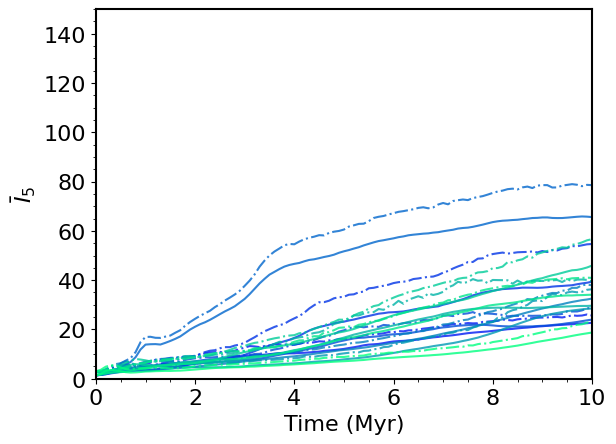}}
 \hspace{0.8pt}
 \subfigure[supervirial (x,z)]{\includegraphics[width=0.31\linewidth]{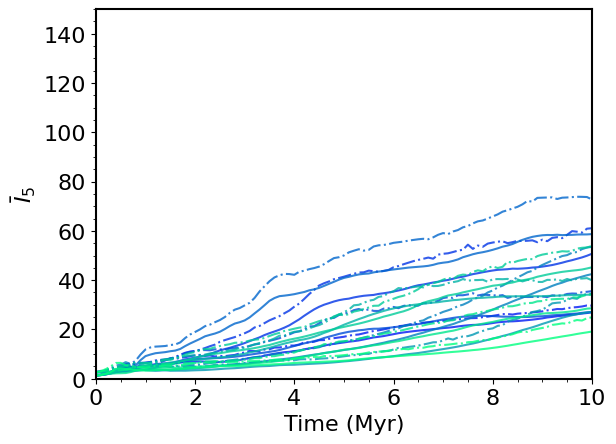}}
 \hspace{0.8pt}
 \subfigure[supervirial (y,z)]{\includegraphics[width=0.31\linewidth]{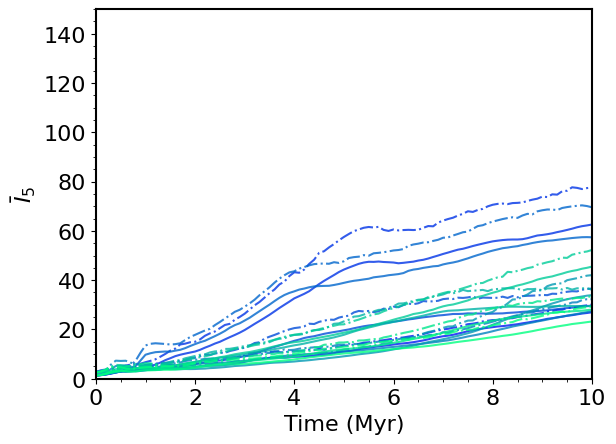}}
    
    \caption{The mean INDICATE index against time for smooth, high-density, sub- and supervirial simulations.  Both the sub- and supervirial simulations have no initial substructure ($D_{f} = 3.0$) and both contain 1000 stars with initial radii of 1 pc, simulated for 10 Myr. From left to right the columns show the INDICATE indexes calculated for three different lines of sight, $(x,y)$, $(x,z)$ and $(y,z)$. The top row shows the indexes calculated for subvirial simulations and the bottom row shows the indexes calculated for supervirial simulations. The solid lines show the mean index for all stars and the dashed lines show the mean index for the significantly clustered stars. We show the 10 simulations we run for the different sets of initial conditions, with the colour shading of the lines being consistent across the different lines of sight.}
    \label{fig:3d0 1pc}
\end{figure*}

Figure~\ref{fig:3d0 1pc} shows the mean INDICATE indexes against time for regions with no initial substructure  ($D_{f} = 3.0$) and initial radii of 1 pc. When comparing between the different virial states we see similar behaviour as for sub- and supervirial regions with $D_{f} = 1.6$. We find that the final mean indexes are similar to those of the regions with $D_{f} = 1.6$. 

For subvirial regions (Fig.~\ref{fig:3d0 1pc}, panels a--c) we see once again an initial rapid increase in the clustering before a gradual increase for the remaining duration of the simulations. The initial increase lasts longer in these simulations, with the increase lasting for around 3 Myr, compared to it finishing within 2 Myr for the substructured ($D_{f}=1.6$) simulations. We see similar behaviour for the supervirial simulations (Fig.~\ref{fig:3d0 1pc}, panels d--f) as well, in that they do not attain as high a mean INDICATE index by the end of the simulations.

\begin{figure*}
 \subfigure[subvirial (x,y)]{\includegraphics[width=0.31\linewidth]{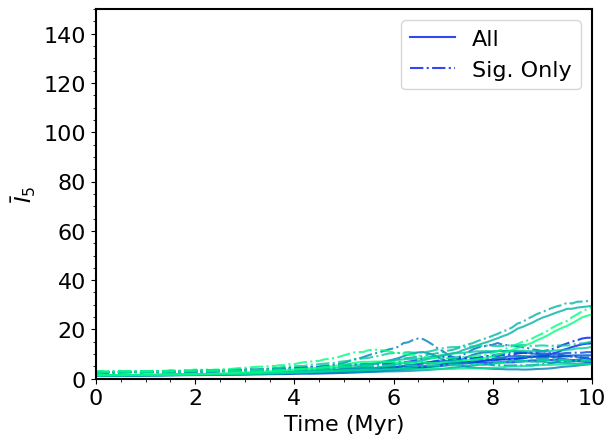}}
 \hspace{0.8pt}
 \subfigure[subvirial (x,z)]{\includegraphics[width=0.31\linewidth]{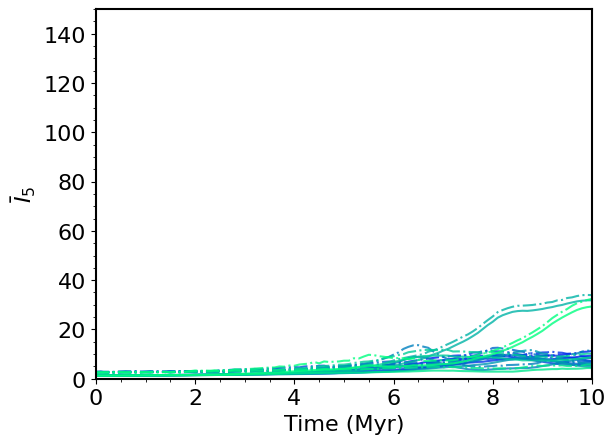}}
 \hspace{0.8pt}
 \subfigure[subvirial (y,z)]{\includegraphics[width=0.31\linewidth]{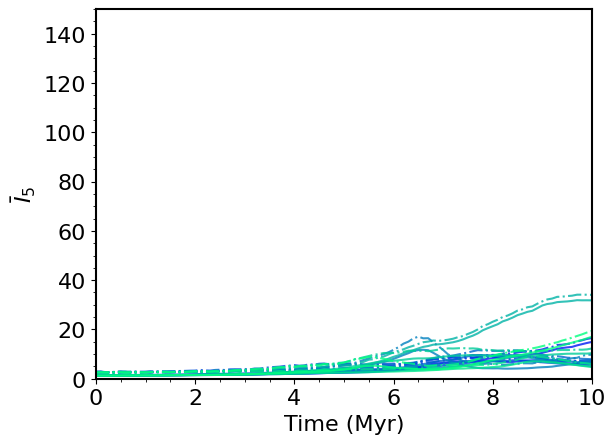}}
 \hspace{0.8pt}
 \subfigure[supervirial (x,y)]{\includegraphics[width=0.31\linewidth]{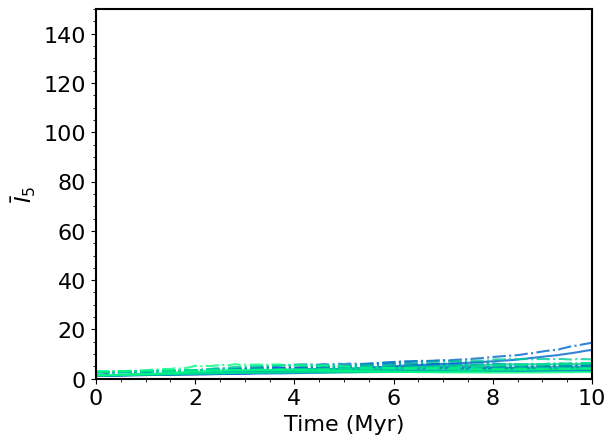}}
 \hspace{0.8pt}
 \subfigure[supervirial (x,z)]{\includegraphics[width=0.31\linewidth]{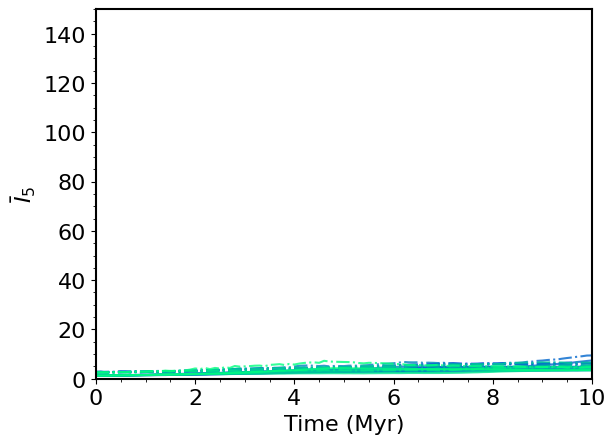}}
 \hspace{0.8pt}
 \subfigure[supervirial (y,z)]{\includegraphics[width=0.31\linewidth]{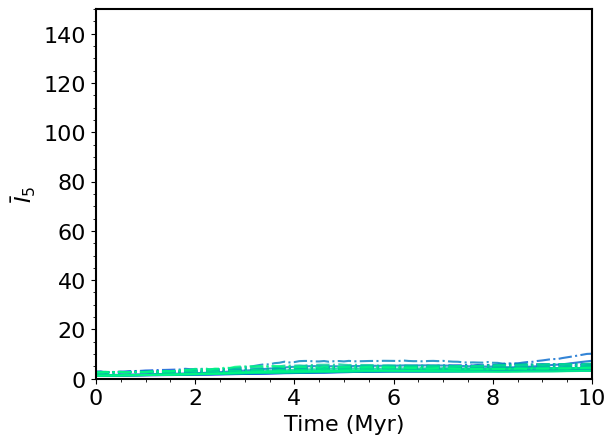}}
    
    \caption{The mean INDICATE index against time for smooth, low-density, sub- and supervirial simulations.  Both the sub- and supervirial simulations have no initial substructure ($D_{f} = 3.0$) and both contain 1000 stars with initial radii of 5 pc, simulated for 10 Myr. From left to right the columns show the INDICATE indexes calculated for different viewing angles, $(x,y)$, $(x,z)$ and $(y,z)$. The top row shows the indexes calculated for subvirial simulations and the bottom row shows the indexes calculated for supervirial simulations. The solid lines show the mean index for all stars and the dashed lines show the mean index for the significantly clustered stars. We show the 10 simulations we run for the different sets of initial conditions, with the colour shading of the lines being consistent across the different lines of sight.}
    \label{fig:3d0 5pc}
\end{figure*}

Figure~\ref{fig:3d0 5pc} shows $\bar{I}_{5}$ against time for low density regions with no substructure ($D_{f} = 3.0$) and initial radii of 5 pc. We see no significant difference in the evolution of the clustering between the different lines of sight. For subvirial simulations (Fig.~\ref{fig:3d0 5pc}, panels a--c) there is no rapid increase in $\bar{I}_{5}$ as the stars are further apart initially, and the absence of substructure also prevents the development of clustering.  This delay is substantial; there is an increase in the clustering for some of the simulations near the end of the run time, at around 7 Myr, but the majority of simulations' INDICATE indexes do not increase. For the supervirial simulations (Fig.~\ref{fig:3d0 5pc}, panels d--f) the clustering does not change significantly with $\bar{I}_{5} < 20$ for 10 Myr.

% 3d0 1pc radii

% 3d0 5 pc radii

\subsection{Combining Methods}
In this section we present the 2D results (line of sight along the $z$ axis with the plane of sky being $x$ and $y$) of combining the Q-parameter, $\Sigma_{\rm{LDR}}$ and $\Lambda_{\rm{MSR}}$ with INDICATE. We do this for each of the simulations at the following times: 0 Myr, 1 Myr and 5 Myr (see also \citet{parker_dynamical_2014_mnras}, \citet{parker_comparisons_2015} and \citet{blaylock-squibbs_evolution_2023} for other examples of Q-parameter, $\Sigma_{\rm{LDR}}$ and $\Lambda_{\rm{MSR}}$ being used in combination).

\subsubsection{Q \& INDICATE}
\begin{figure*}
 \subfigure[1 pc, subvirial, high density]{\includegraphics[width=0.49\linewidth]{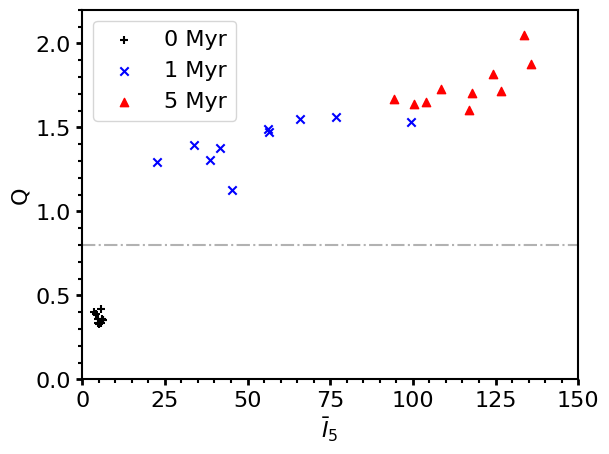}}
 \hspace{0.8pt}
 \subfigure[1 pc, supervirial, high density]{\includegraphics[width=0.49\linewidth]{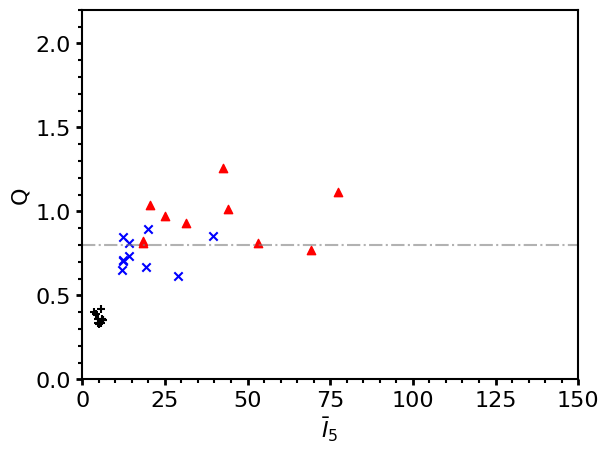}}
 \hspace{0.8pt}
 \subfigure[5 pc, subvirial, low density]{\includegraphics[width=0.49\linewidth]{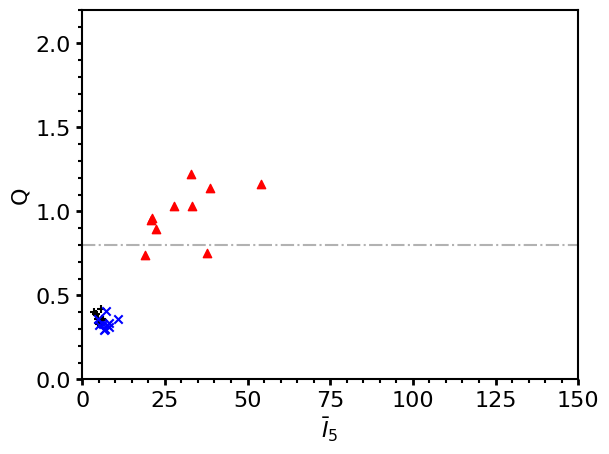}}
 \hspace{0.8pt}
 \subfigure[5 pc, supervirial, low density]{\includegraphics[width=0.49\linewidth]{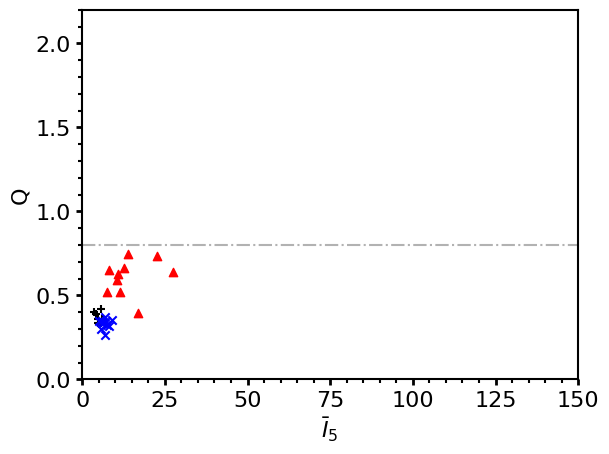}}
    
    \caption{The Q-parameter plotted against INDICATE for simulations with a high degree of initial substructure ($D_{f}=1.6$). The top row shows the results for high density simulations with initial radii of 1 pc and the bottom row shows the results for low density simulations with initial radii of 5 pc. The black pluses show the results at 0 Myr, the blue crosses for 1 Myr and the red triangles are for 5 Myr. The horizontal grey dash dotted line is at 0.8, above which Q values signify the regions have a smooth, centrally concentrated morphology. The left hand column shows the results for subvirial simulations and the right hand column shows the results for supervirial simulations. }
    \label{fig:1d6 indicate vs q}
\end{figure*}

% q vs indicate 1.6
Figure~\ref{fig:1d6 indicate vs q} shows INDICATE plotted against Q at 0 Myr, 1 Myr and 5 Myr in the high density, substructured ($D_{f} = 1.6$) simulations, for subvirial (panel a) and supervirial (panel b) simulations, respectively. Comparing the sub- and supervirial simulations we see a clear distinction between them in both Q and $\bar{I}_{5}$. All the subvirial simulations have $Q > 1$ by 1 Myr, and at the same time $\bar{I}_{5}$ lies between 20 and 100. The $\bar{I}_{5}$ values found at 5 Myr in the subvirial simulations overlap slightly with the 1 Myr results, lying between 85 and 135. 

In the high density supervirial simulations (Figure~\ref{fig:1d6 indicate vs q}(b)) we find that some of the Q values at 1 Myr and 5 Myr have remained below the boundary between substructured and smooth morphologies (stars are moving away from each other in grouping of stars and this preserves some of the initial substructure, hence a lower Q value), with some Q values at 5 Myr overlapping with the 1 Myr values found in the subvirial simulations. The overlap between the $\bar{I}_{5}$ values at 1 Myr and 5 Myr has also increased, making distinguishing between the different times more difficult.

Panels~(c) and (d) of Figure~\ref{fig:1d6 indicate vs q} show the results for the low density, initially substructured ($D_{f} = 1.6$) simulations. The INDICATE results for 0 Myr and 1 Myr are similar when comparing the sub- and supervirial realisations, due to a lack of dynamical evolution at early times. We once again see that the 5 Myr Q values are higher in the subvirial regions as the substructure is more effectively erased due to the stars collapsing inwards into the gravitational potential well. On the other hand, the supervirial regions retain substructure, but do not become more clustered due to a lack of significant dynamical evolution.%If we compare the INDICATE values at 5 Myr between the different initial virial states there is significant overlap between them, with subvirial values between 15 and 55 and supervirial values between 5 and 30.
% RJP - I don't agree with this last sentence, there's quite good distinction at 5 Myr.

\begin{figure*}
 \subfigure[1 pc, subvirial, high density]{\includegraphics[width=0.49\linewidth]{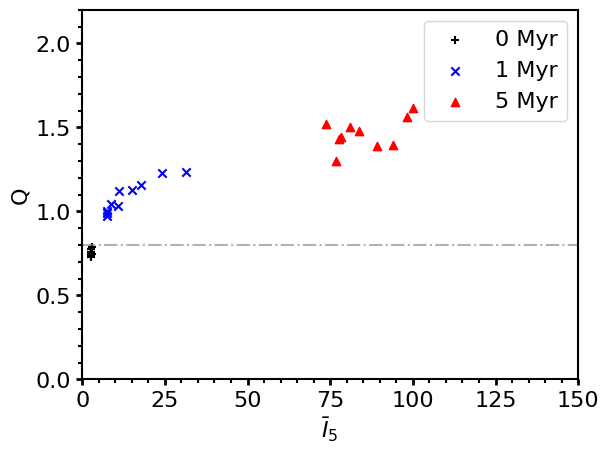}}
 \hspace{0.8pt}
 \subfigure[1 pc, supervirial, high density]{\includegraphics[width=0.49\linewidth]{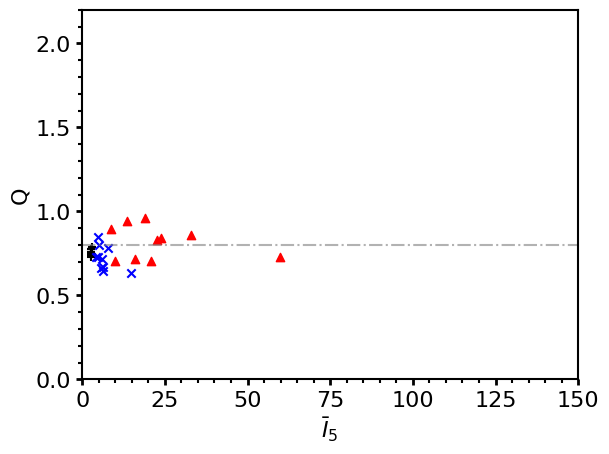}}
 \hspace{0.8pt}
 \subfigure[5 pc, subvirial, low density]{\includegraphics[width=0.49\linewidth]{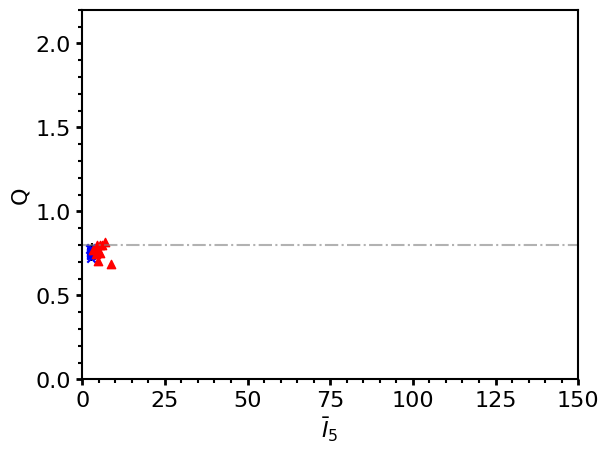}}
 \hspace{0.8pt}
 \subfigure[5 pc, supervirial, low density]{\includegraphics[width=0.49\linewidth]{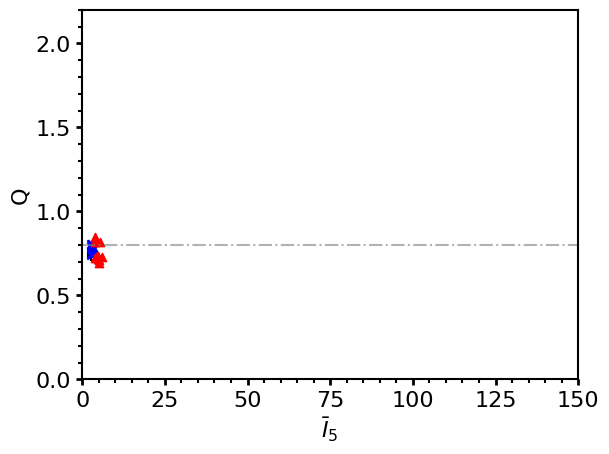}}
    
    \caption{The Q-parameter plotted against INDICATE for simulations with little to no initial substructure ($D_{f}=3.0$). The top row shows the results for high density simulations with initial radii of 1 pc and the bottom row shows the results for low density simulations with initial radii of 5 pc. The black pluses show the results at 0 Myr, the blue crosses for 1 Myr and the red triangles are for 5 Myr. The horizontal grey dash dotted line is at 0.8, above which Q values signify the regions have a smooth, centrally concentrated morphology. The left hand column shows the results for subvirial simulations and the right hand column shows the results for supervirial simulations. The top row shows the high density simulations and the bottom row shows the low density simulations.}
    \label{fig:3d0 indicate vs q}
\end{figure*}

% q vs indicate 3.0
Figure~\ref{fig:3d0 indicate vs q} shows the Q and $\bar{I}_{5}$ values for initially smooth subvirial (lefthand panels) and supervirial (righthand panels) simulations, respectively. Panels (a) and (b) show simulations that are high density with $D_{f} = 3.0$ (little to no initial degree of substructure). For subvirial simulations (panel a) both Q and $\bar{I}_{5}$ values are distinct with very little overlap between the different times. For  the supervirial simulations (panel b) there is much more overlap, with Q values being similar across 0 Myr, 1 Myr and 5 Myr, and only the INDICATE vales increase slightly over time. As in the highly substructured simulations, we find that the maximum $\bar{I}_{5}$ values are found in the subvirial simulations.

In panels~(c) and (d) of Figure~\ref{fig:3d0 indicate vs q} we show the evolution of non-substructured ($D_{f} = 3.0$), low-density ($r = 5$\,pc) simulations and we find the Q values overlap across 0 Myr, 1 Myr and 5 Myr. This overlap is seen for both subvirial and supervirial simulations implying that the initial virial state of low density regions without substructure cannot be constrained using these methods.

\subsubsection{$\Sigma_{\rm{LDR}}$ \& INDICATE}
% 1.6 1pc 
\citet{parker_dynamical_2014_mnras} and \citet{parker_dynamics_2014} show that the relative surface densities of the most massive stars can be used in tandem with the Q-parameter to infer the initial density and virial state of a star-forming region. 

\begin{figure*}
 \subfigure[1 pc, subvirial, high density]{\includegraphics[width=0.49\linewidth]{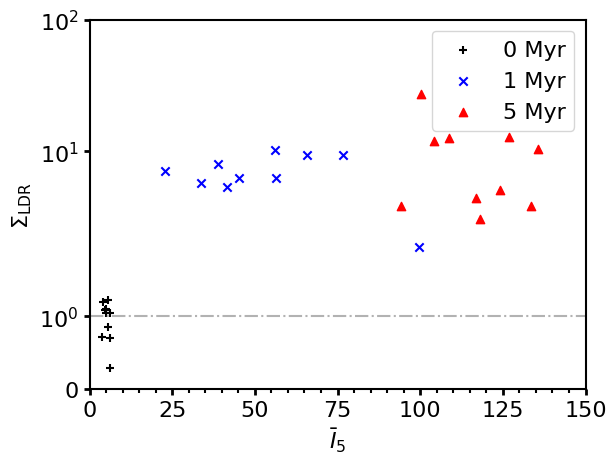}}
 \hspace{0.8pt}
 \subfigure[1 pc, supervirial, high density]{\includegraphics[width=0.49\linewidth]{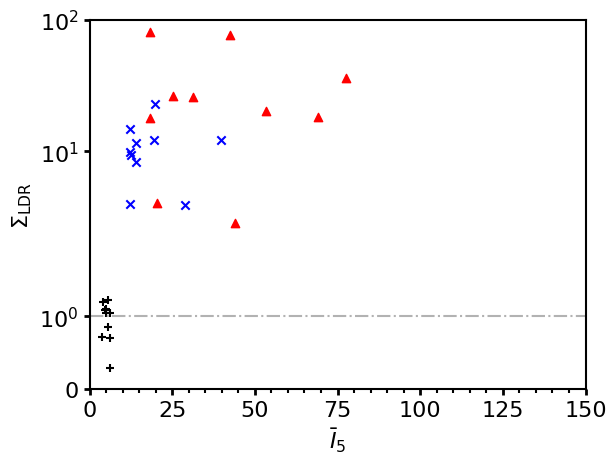}}
 \hspace{0.8pt}
 \subfigure[5 pc, subvirial, low density]{\includegraphics[width=0.49\linewidth]{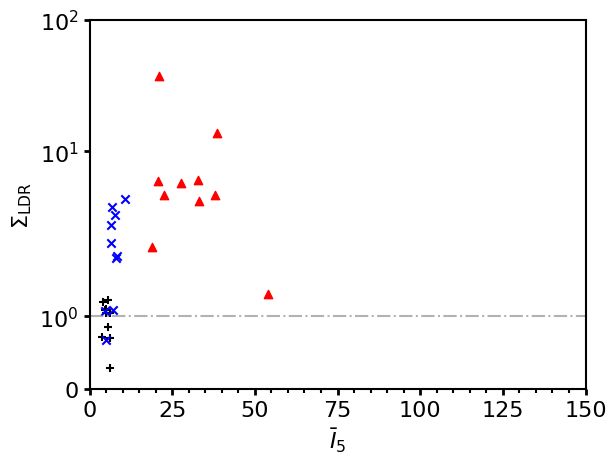}}
 \hspace{0.8pt}
 \subfigure[5 pc, supervirial, low density]{\includegraphics[width=0.49\linewidth]{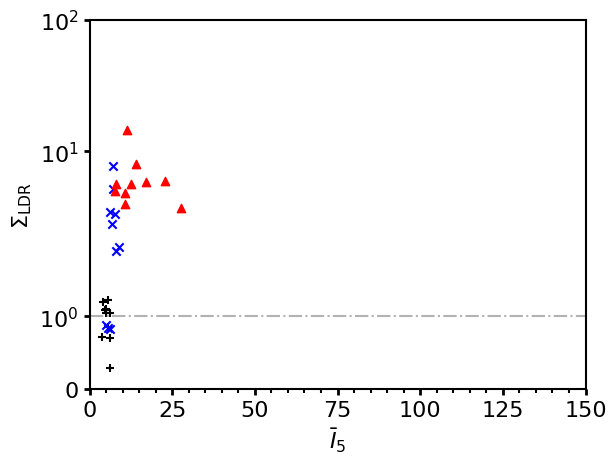}}
    
    \caption{The relative surface density of massive stars, $\Sigma_{\rm{LDR}}$, plotted against INDICATE for simulations with a high degree of initial substructure ($D_{f}=1.6$). The black pluses show the results for 0 Myr, the blue crosses show the results for 1 Myr and the red triangles for 5 Myr. The horizontal grey dash dotted line is at 1, above which $\Sigma_{\rm{LDR}}$ finds the 10 most massive stars are in areas of greater than average surface density. The top row shows the results for high-density simulations with initial radii of 1 pc and the bottom row shows the results for low-density simulations with initial radii of 5 pc. The left hand column shows the results for subvirial simulations and the right hand column shows the results for the supervirial simulations.}
    \label{fig:1d6 indicate vs sigma}
\end{figure*}

Figures~\ref{fig:1d6 indicate vs sigma}~(a) and (b) show the results for $\Sigma_{\rm{LDR}}$ and INDICATE for high density and highly substructured regions with $D_{f} = 1.6$. The $\Sigma_{\rm{LDR}}$ values are less reliable at inferring the initial conditions than the Q values, with an overlap between $\Sigma_{\rm{LDR}}$ between the sub- and supervirial simulations at 1 Myr and 5 Myr. The $\bar{I}_{5}$ values are the same here so the analysis in the previous section will apply for the rest of this section.

Panels~(c) and (d) show the combination for low density simulations. Comparing across the sub- and supervirial simulations we see some overlap in the $\Sigma_{\rm{LDR}}$ values for the 0 Myr and 1 Myr snapshots, although after 5\,Myr it is possible to distinguish between the high- and low-density simulations. %Comparing between the virial states there is significant overlap in $\Sigma_{\rm{LDR}}$ across all three times. 

\begin{figure*}
 \subfigure[1 pc, subvirial, high density]{\includegraphics[width=0.49\linewidth]{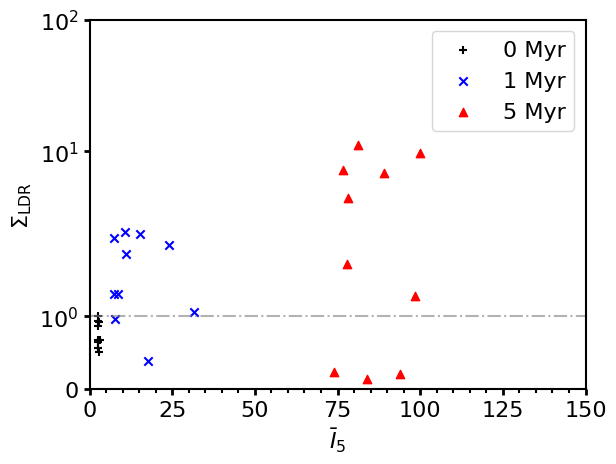}}
 \hspace{0.8pt}
 \subfigure[1 pc, supervirial, high density]{\includegraphics[width=0.49\linewidth]{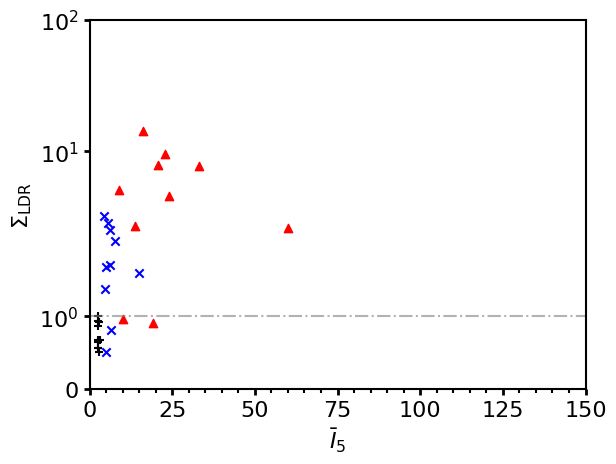}}
 \hspace{0.8pt}
 \subfigure[5 pc, subvirial, low density]{\includegraphics[width=0.49\linewidth]{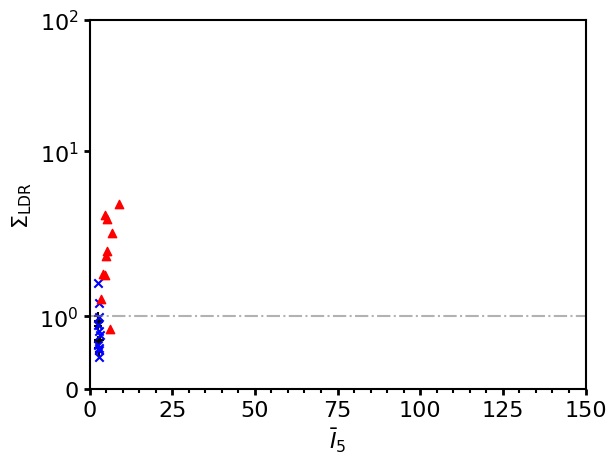}}
 \hspace{0.8pt}
 \subfigure[5 pc, supervirial, low density]{\includegraphics[width=0.49\linewidth]{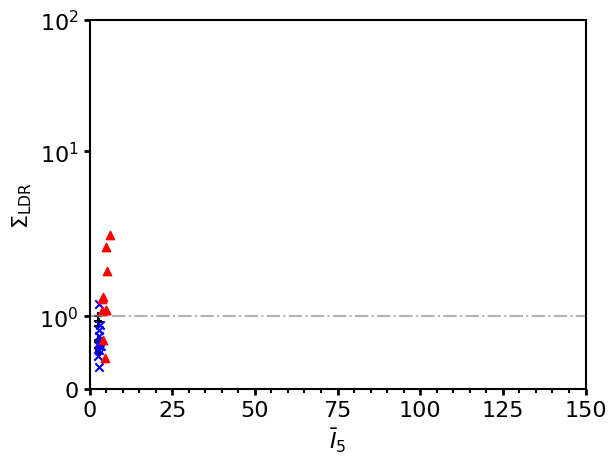}}
    
    \caption{The relative surface density of massive stars, $\Sigma_{\rm{LDR}}$, plotted against INDICATE for simulations with no initial substructure ($D_{f}=3.0$). The black pluses show the results for 0 Myr, the blue crosses show the results for 1 Myr and the red triangles for 5 Myr. The horizontal grey dash dotted line is at 1, above which $\Sigma_{\rm{LDR}}$ finds the 10 most massive stars are in areas of greater than average surface density. The top row shows the results for high-density simulations with initial radii of 1 pc and the bottom row shows the results for low-density simulations with initial radii of 5 pc. The left hand column shows the results for subvirial simulations and the right hand column shows the results for the supervirial simulations.}
    \label{fig:3d0 indicate vs sigma}
\end{figure*}

Figures~\ref{fig:3d0 indicate vs sigma}~(a) and (b) shows the results for dense simulations with no initial substructure ($D_{f} = 3.0$). The dense, subvirial simulations (panel a) show a very distinct difference between 1\,Myr and 5\,Myr (compared the triangles with the crosses), but the distinction is less clear for the supervirial simulations (ostensibly because the degree of clustering measured by INDICATE does not increase as much in the supervirial simulations). 

In the low-density simulations without substructure (panels~(c) and (d) of Figure~\ref{fig:3d0 indicate vs sigma}) there is significant overlap of $\Sigma_{\rm{LDR}}$ versus $\bar{I}_{5}$ at all snapshot times, and as such it is impossible to differentiate between initially subvirial and supervirial simulations. 

% sigma indicate compare the different fractal dimensions.
We find that in the substructured simulations $\Sigma_{\rm{LDR}}$ values tend to be higher at 5 Myr compared to simulations with no substructure, and the degree of clustering measured by INDICATE is also slightly higher in the substructured simulations.% Comparing across equivalent simulation sets (same initial degree of substructure, density and virial state) we find overlap between the $\Sigma_{\rm{LDR}}$ values. Meaning that $\Sigma_{\rm{LDR}}$ cannot reliably infer the initial degree of substructure in a region.

\subsubsection{$\Lambda_{\rm{MSR}}$ \& INDICATE}

In addition to plotting the evolution of the Q parameter against the relative surface density ratio, $\Sigma_{\rm LDR}$, \citet{parker_dynamical_2014_mnras} showed that it is possible to also distinguish between sub- and supervirial initial conditions by using a combination of Q and the mass segregation ratio, $\Lambda_{\rm MSR}$.

\begin{figure*}
 \subfigure[1 pc, subvirial, high density]{\includegraphics[width=0.49\linewidth]{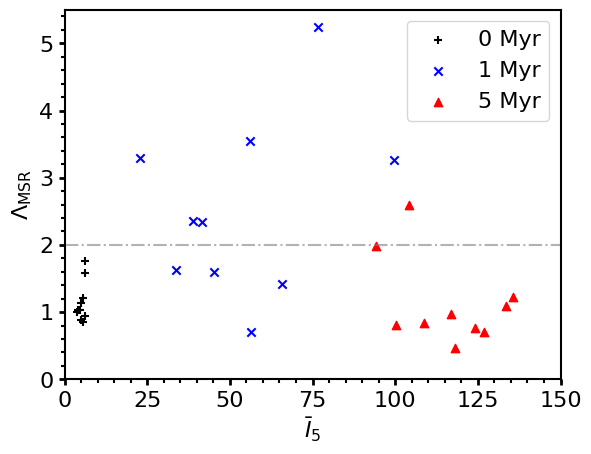}}
 \hspace{0.8pt}
 \subfigure[1 pc, supervirial, high density]{\includegraphics[width=0.49\linewidth]{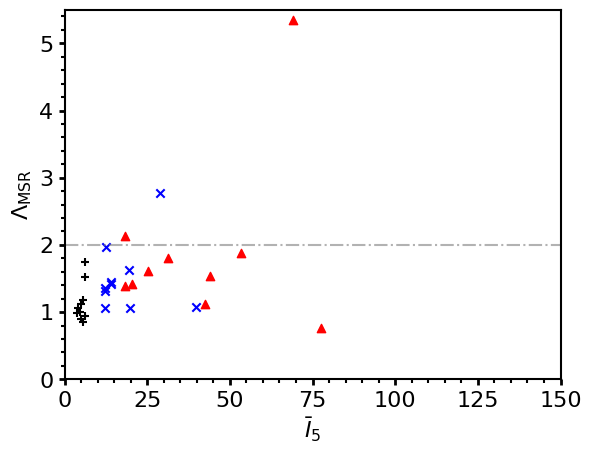}}
 \hspace{0.8pt}
 \subfigure[5 pc, subvirial, low density]{\includegraphics[width=0.49\linewidth]{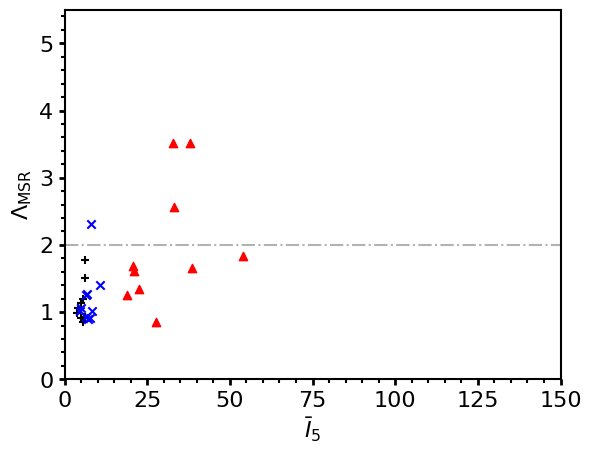}}
 \hspace{0.8pt}
 \subfigure[5 pc, supervirial, low density]{\includegraphics[width=0.49\linewidth]{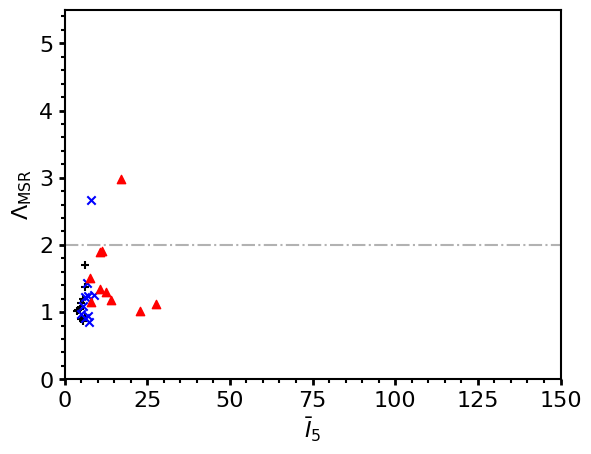}}
    
    \caption{$\Lambda_{\rm{MSR}}$ versus INDICATE for simulations with a high degree of initial substructure ($D_{f}=1.6$). The black pluses show the results for 0 Myr, the blue crosses show for 1 Myr and the red triangles for 5 Myr. The grey dash dotted line is at $\Lambda_{\rm{MSR}} = 2$, above this the 10 most massive stars are mass segregated. The top row shows the results for high density simulations with initial radii of 1 pc and the bottom row shows the results for low density simulations with initial radii of 5 pc. The left hand column shows the results for subvirial simulations and the right hand column shows results for supervirial simulations.}
    \label{fig:1d6 indicate vs lambda}
\end{figure*}

Figure~\ref{fig:1d6 indicate vs lambda} shows the combination of $\Lambda_{\rm{MSR}}$ and INDICATE for initially substructured simulations ($D_f = 1.6$). Although subvirial simulations tend to attain higher $\Lambda_{\rm MSR}$ values \citep{allison_early_2010,parker_dynamical_2014_mnras}, not all simulations dynamically segregate. This, combined with the wide spread in INDICATE indexes makes inferring the initial virial state of dense regions using $\Lambda_{\rm{MSR}}$ and INDICATE alone unreliable (compare  panels~(a) and (b)). The $\Lambda_{\rm{MSR}}$ values are also similar between the subvirial and supervirial simulations in the low density initial conditions (panels c and d). The key difference between the low and high density simulations is the $\Lambda_{\rm MSR}$ values at 1 Myr tending to be higher in the subvirial high density simulations compared to the supervirial high density simulations. 

\begin{figure*}
 \subfigure[1 pc, subvirial, high density]{\includegraphics[width=0.49\linewidth]{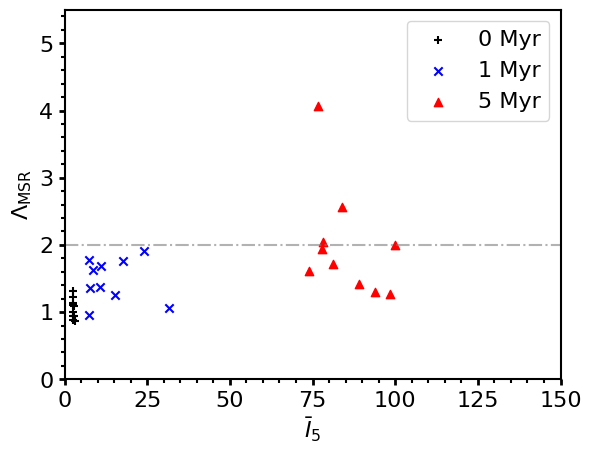}}
 \hspace{0.8pt}
 \subfigure[1 pc, supervirial, high density]{\includegraphics[width=0.49\linewidth]{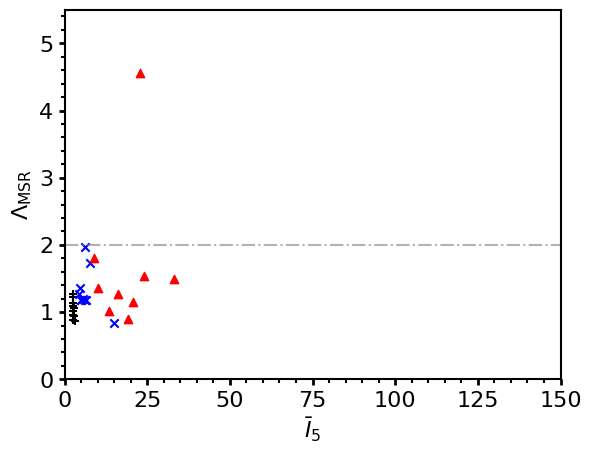}}
 \hspace{0.8pt}
 \subfigure[5 pc, subvirial, low density]{\includegraphics[width=0.49\linewidth]{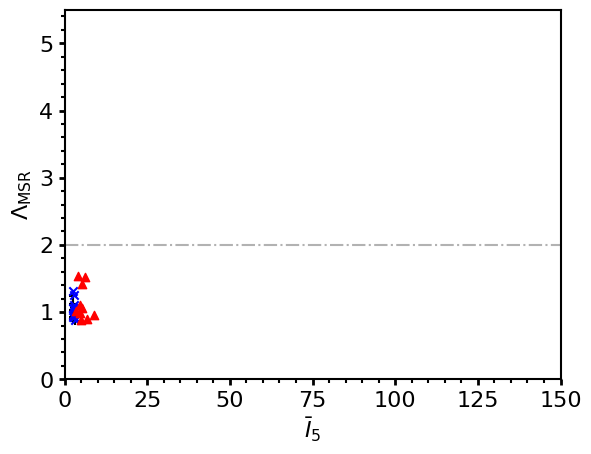}}
 \hspace{0.8pt}
 \subfigure[5 pc, supervirial, low density]{\includegraphics[width=0.49\linewidth]{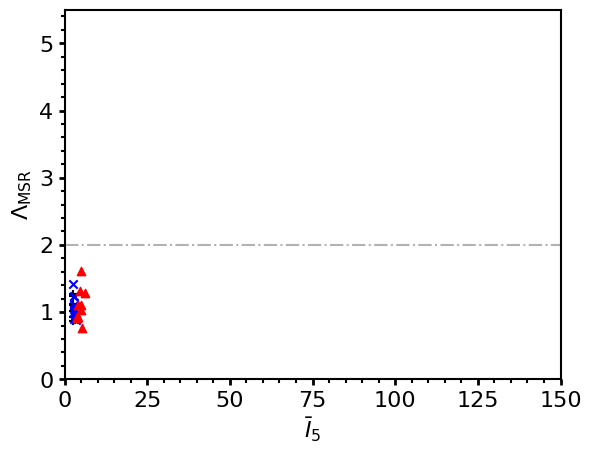}}
    
    \caption{$\Lambda_{\rm{MSR}}$ versus INDICATE for simulations with no  initial substructure ($D_{f}=3.0$). The black pluses show the results for 0 Myr, the blue crosses show for 1 Myr and the red triangles for 5 Myr. The grey dash dotted line is at $\Lambda_{\rm{MSR}} = 2$, above this the 10 most massive stars are mass segregated. The top row shows the results for high density simulations with initial radii of 1 pc and the bottom row shows the results for low density simulations with initial radii of 5 pc. The left hand column shows the results for subvirial simulations and the right hand columns shows the results for supervirial simulations.}
    \label{fig:3d0 indicate vs lambda}
\end{figure*}

Figure~\ref{fig:3d0 indicate vs lambda} shows the results for simulations with no initial substructure ($D_{f} = 3.0$). Comparing the high density simulations we see no significant difference in the $\Lambda_{\rm{MSR}}$ values, with a single simulation in both sets attaining $\Lambda_{\rm{MSR}} > 4$, which is a significant degree of mass segregation.

However, the combination of $\Lambda_{\rm MSR}$ and INDICATE does provide some constraints; the smooth, subvirial simulations (panel a) are distinguishable from the supervirial simulations (panel b). 

For the low-density simulations (panels c and d) the two different initial virial states are indistinguishable.

%Comparing the low density simulations in panels~(c) and (d) there is significant overlap between values at 0 Myr, 1 Myr and 5 Myr. 

% comparing the simulations with different degrees of substructure 
%Comparing the simulations with and without substructure we find that the $\Lambda_{\rm{MSR}}$ values overlap in all sets of simulations.

\subsection{Determining the initial degree of substructure}

We have shown that combining INDICATE with other measures of the spatial distribution of stars can be used to infer the initial density of a star-forming region, and whether the star-forming region was initially subvirial, or supervirial. 

Observations suggest that many star-forming regions form with the stars arranged in a substructured distribution, likely to be inherited from the filamentary nature of the gas from which they form. Substructure is not created during the dynamical evolution of a star-forming region, it is always erased \citep{goodwin_dynamical_2004,parker_dynamical_2014_mnras,daffern-powell_dynamical_2020}. 

Hydrodynamical simulations of the early phases of star formation suggest that the initial degree of substructure of stars can vary quite considerably \citep{schmeja_evolving_2006,girichidis2012,dale2012,dale2013,dale2014}. As any observed star-forming region could have experienced some dynamical evolution, it is not clear what the typical degree of substructure is for star-forming regions (or even if there is a typical value).  

Furthermore, \citet{dib_and_henning_sf_spatial_distribution_ms_2019} suggest that the value of the Q-parameter increases with increasing star formation rate, and postulate a link between the early stages of star formation in clouds, and the later degree of substructure. This could, however, be affected by any later dynamical evolution of the stars \citep{parker_dynamical_2014_mnras, parker_no_2017}, and if there is not a direct mapping of the structure of the gas to the stars \citep{parker_spatial_2015}.

Inspection of our results suggests that combining INDICATE with other measures of the spatial (and kinematic) information could be used to infer the initial degree of substructure. The most important factor in determining the amount of dynamical evolution is the initial local density within the star-forming region \citep{parker_dynamics_2014}. This means that in order to compare the effects of the substructure on the evolution of a star-forming region, we must compare regions with similar median local densities. 

For regions with similar radii, a highly substructured region ($D_{f}=1.6$) will have a very high median density compared to a region with no substructure ($D_{f}=3.0$). Our substructured regions with large radii (5\,pc) have similar median local densities ($\tilde{\rho} \sim 100\,$M$_\odot$\,pc$^{-3}$) to our smooth regions ($D_{f}=3.0$) with smaller radii (1\,pc), and so we determine the effects of substructure on the evolution of INDICATE using these pairs of simulations.

As an example, we compare the evolution of the Q-parameter and INDICATE for substructured simulations with large radii (Fig.~\ref{fig:1d6 indicate vs q}-c) to smooth simulations with more compact radii (Fig.~\ref{fig:3d0 indicate vs q}-a). We can see that both Q and INDICATE display a higher level of clustering in the more compact, smoother simulations, and this is also evident in the subsequent plots that show the evolution of INDICATE combined with $\Lambda_{\rm MSR}$ (compare Fig.~\ref{fig:1d6 indicate vs lambda}-c to Fig.~\ref{fig:3d0 indicate vs lambda}-a) or $\Sigma_{\rm LDR}$ (compare Fig.~\ref{fig:1d6 indicate vs sigma}-c to Fig.~\ref{fig:3d0 indicate vs sigma}-a).

This is interesting because for most combinations of metrics, their signal is enhanced from the dynamical evolution of initially substructured regions. Because the degree of clustering measured by INDICATE is highest for smooth rather than substructured regions (of comparable median local densities), INDICATE can be used in tandem with other metrics to distinguish between subtle differences in the initial degree of substructure in star-forming regions. 

\section{Conclusions}
We apply the INDICATE clustering measure \citep{buckner_spatial_2019} to eight sets of $N$-body simulations consisting of 1000 stars along three different lines of sight to understand how it behaves due to dynamical evolution in star-forming regions with different initial conditions. We also combine INDICATE with other methods to asses its diagnostic ability in determining the initial conditions of the simulations from later snapshots. Our conclusions are as follows.
\begin{enumerate}
    \item INDICATE is not significantly affected by projection affects in our simulations, and evolves similarly when measured along different lines of sight for the same set of simulations.
    
    \item INDICATE does evolve differently depending on the initial virial state of the simulations. We find that $\bar{I}_5$ increases rapidly in the subvirial simulations during the first 2 Myr, and then steadily increases until the end of the simulations. This behaviour is seen in both simulations with a high degree of initial substructure ($D_{f} = 1.6$) and in simulations with no initial substructure ($D_{f} = 3.0$), albeit to a lesser extent in the less substructured simulations.  

       \item The values of the INDICATE measure, $\bar{I}_{5}$, are higher for subvirial simulations than for supervirial simulations at the end of the simulations (10\,Myr), and this is seen in all sets of simulations apart from very low density, initially non-substructured $D_{f} = 3.0$ simulations, where there has been little-to-no dynamical evolution.

    \item INDICATE is most sensitive to the initial density, rather than the initial degree of substructure in a region. The higher the initial density of a star-forming region, the more clustered the stars tend to become.

    \item However, INDICATE -- in combination with other metrics such as the Q-parameter -- can be used to infer the initial degree of substructure, but only if the initial densities and virial ratios are known. For example, compare Fig.~\ref{fig:1d6 indicate vs q} and Fig.~\ref{fig:3d0 indicate vs q}, where the plot of Q versus $\bar{I}_{5}$ gives different results for high density simulations with and without substructure. This difference is much less pronounced for the low density simulations.
 
    \item When we combine INDICATE with other measures of spatial clustering, we find that INDICATE used in combination with the Q-parameter \citep{cartwright_statistical_2004} provides the most reliable way of inferring the initial conditions of the simulated star-forming regions.

    %\item \textbf{We find that both Q and INDICATE give similar results for low density regions that have different initial degree of substructure, meaning that the initial virial state or degree of substructure cannot be inferred reliably.}

    %\item \textbf{We find that the Q-parameter is more reliable at inferring the initial conditions of the simulations, more so than INDICATE. This is most evident when comparing the low density regions with different degrees of substructure in Figures~\ref{fig:1d6 indicate vs q} and \ref{fig:3d0 indicate vs q}. Low density regions with a high initial degree of substructure attain higher Q values than low density regions without substructure. This is because in the substructured regions more dynamical processing tends to take place, erasing substructure and increasing the Q value. There is a pronounced difference in the Q values when comparing the 0 Myr and 1 Myr results in which substructured regions have Q$\, < 0.5$ and regions with no substructure have Q$\, > 0.5$.}
    
\end{enumerate}

Ideally, combinations of more than one spatial diagnostic (e.g. INDICATE and Q, Q and $\Sigma_{\rm LDR}$, etc.) are required to really pinpoint the initial conditions of a star-forming region, in addition to kinematic measures such as the distributions of ejected stars \citep{schoettler_dynamical_2019,schoettler_runaway_2020,schoettler2022,farias2020,parker2022,arunima2023}. 

\section*{Acknowledgements}
The plots in this paper have been created using Matplotlib 3.3.4 \citep{hunter_matplotlib_2007}. Numerical results have been calculated using Numpy and SciPy 1.9.0 \citep{harris_array_2020, virtanen_scipy_2020}.

GABS acknowledges support from the University of Sheffield in the form of a publication scholarship and the University of Sheffield Institutional Open Access Fund. RJP acknowledges support from the Royal
Society in the form of a Dorothy Hodgkin Fellowship.

For the purpose of open access, the author has applied a Creative Commons Attribution (CC BY) licence to any Author Accepted Manuscript version arising from this submission.
\clearpage
% We wish to thank the anonymous referee for their feedback which has improved this work.
%%%%%%%%%%%%%%%%%%%%%%%%%%%%%%%%%%%%%%%%%%%%%%%%%%
\section*{Data Availability}
% The data used to generate the plots in this paper will be provided upon reasonable request to either of the corresponding authors.

The simulation data used in this paper is available publicly on the ORDA data repository. Subvirial simulation set: \href{https://orda.shef.ac.uk/articles/dataset/Subvirial_Simulations/25205207}{\url{https://orda.shef.ac.uk/articles/dataset/Subvirial_Simulations/25205207}}. Supervirial simulation set: \href{https://orda.shef.ac.uk/articles/dataset/Supervirial_Simulations/25205846}{\url{https://orda.shef.ac.uk/articles/dataset/Supervirial_Simulations/25205846}}.

\bibliographystyle{mnras}
\bibliography{bibliography_iot} % if your bibtex file is called example.bib

%%%%%%%%%%%%%%%%%%%%%%%%%%%%%%%%%%%%%%%%%%%%%%%%%%

%%%%%%%%%%%%%%%%% APPENDICES %%%%%%%%%%%%%%%%%%%%%

% \appendix

% \section{Some extra material}

%%%%%%%%%%%%%%%%%%%%%%%%%%%%%%%%%%%%%%%%%%%%%%%%%%

% Don't change these lines
\bsp	% typesetting comment
\label{lastpage}
\end{document}